\documentclass[12pt]{article}
\usepackage{amsmath,amssymb,bm}
\usepackage{graphicx}
\usepackage{enumerate}
\usepackage{natbib}
\usepackage[colorlinks=true,allcolors=blue]{hyperref}%
\usepackage{url} 
\usepackage[percent]{overpic}
\usepackage{caption}
\usepackage{graphicx,subfigure}
\usepackage{dsfont}
\usepackage{algorithm,algorithmic}
\usepackage[normalem]{ulem}
\usepackage{xcolor}
\usepackage{epstopdf,multirow,lscape,comment}
\usepackage{arydshln} 

\newcommand{\blind}{0}

\newcommand{\qed}{$\hfill\blacksquare$}
\newtheorem{Th}{{\bf Theorem}}

\newtheorem{Rem}{{\bf Remark}}

\newtheorem{proposition}{Proposition}
\newtheorem{Lem}{{\bf Lemma}}
\newtheorem{Cor}{{\bf Corollary}}

\newtheorem{Ass}{Assumption}

\def\bA{{\mathbf A}}
\def\ba{{\mathbf a}}

\def\bC{{\mathbf C}}
\def\bD{{\mathbf D}}
\def\bd{{\mathbf d}}
\def\bE{{\mathbf E}}
\def\b1e{{\mathbf e}}

\def\bG{{\mathbf G}}

\def\bI{{\mathbf I}}

\def\bP{{\mathbf P}}

\def\bqu{{\mathbf q}}

\def\bS{{\mathbf S}}
\def\bs{{\mathbf s}}

\def\by{{\mathbf y}}
\def\bY{{\mathbf Y}}

\def\bz{{\mathbf z}}

\def\bx{{\mathbf x}}
\def\bX{{\mathbf X}}

\def\bV{{\mathbf V}}

\def\bzero{{\mathbf 0}}

\def\bnu{{\boldsymbol{\nu}}}
\def\bleta{{\boldsymbol{\eta}}}
\def\bepsilon{{\boldsymbol{\epsilon}}}

\def\btheta{{\boldsymbol{\theta}}}
\def\bTheta{{\boldsymbol{\Theta}}}

\def\boxit#1{\vbox{\hrule\hbox{\vrule\kern6pt
          \vbox{\kern6pt#1\kern6pt}\kern6pt\vrule}\hrule}}

\def\wt{\widetilde}

\def\wh{\widehat}
\DeclareFontFamily{U}{mathx}{}
\DeclareFontShape{U}{mathx}{m}{n}{<-> mathx10}{}
\DeclareSymbolFont{mathx}{U}{mathx}{m}{n}
\DeclareMathAccent{\widecheck}{0}{mathx}{"71}

\def\argmin{\hbox{argmin}}

\def\Normal{\mathbb{N}}

\def\bse{\begin{eqnarray*}}
\def\ese{\end{eqnarray*}}
\def\be{\begin{eqnarray}}
\def\ee{\end{eqnarray}}
\def\bq{\begin{equation}}
\def\eq{\end{equation}}
\def\bse{\begin{eqnarray*}}
\def\ese{\end{eqnarray*}}

\def\wh{\widehat}

\def\b1e{\boldsymbol{ e}}

\def\bx{{\mathbf x}}
\def\bX{{\mathbf X}}

\def\bu{{\mathbf u}}
\def\bS{{\mathbf S}}

\def\bzero{{\mathbf 0}}

\newcommand*{\mydot}{\mathrel{\scalebox{0.4}{$\bullet$}}}

\newcommand{\bSigma}{\mbox{\boldmath $\Sigma$}}
\newcommand{\bsigma}{\mbox{\boldmath $\sigma$}}

\newcommand{\balpha}{\mbox{\boldmath $\alpha$}}
\newcommand{\bOmega}{\mbox{\boldmath $\Omega$}}

\newcommand{\bmu}{\mbox{\boldmath $\mu$}}

\def\rkcomment#1{\vskip 2mm\boxit{\vskip 2mm{\color{blue}\bf#1} {\color{blue}\bf -- RK\vskip 2mm}}\vskip 2mm}
\newcommand{\rkc}[1]{{\color{red}[RK: #1]}}

\def\rkcomment#1{\vskip 2mm\boxit{\vskip 2mm{\color{blue}\bf#1} {\color{blue}\bf -- RK\vskip 2mm}}\vskip 2mm}

\begin{document}

\def\spacingset#1{\renewcommand{\baselinestretch}%
{#1}\small\normalsize} \spacingset{1}


\if0\blind
{
  \title{A Sparse Linear Model for Positive Definite Estimation of Covariance Matrices}
  \author{Rakheon Kim\\
    Department of Statistical Science, Baylor University\\
    and \\
    Irina Gaynanova \\
    Department of Biostatistics, University of Michigan
    \date{}
    }
  \maketitle
} \fi

\if1\blind
{
  \bigskip
  \bigskip
  \bigskip
  \begin{center}
    {\LARGE\bf Title}
\end{center}
  \medskip
} \fi

\bigskip
\begin{abstract}
Sparse covariance matrices play crucial roles by encoding the interdependencies between variables in numerous fields such as genetics and neuroscience. Despite substantial studies on sparse covariance matrices, existing methods face several challenges such as the correlation among the elements in the sample covariance matrix, positive definiteness and unbiased estimation of the diagonal elements. To address these challenges, we formulate a linear covariance model for estimating sparse covariance matrices and propose a penalized regression. This method is general enough to encompass existing sparse covariance estimators and can additionally consider correlation among the elements in the sample covariance matrix while avoiding unnecessary bias in the diagonal elements and preserving positive definiteness. We develop a consensus ADMM algorithm for estimation and derive $\ell_2$ convergence rate of the proposed estimator. We apply our estimator to simulated data and real data from neuroscience and genetics to describe the efficacy of our proposed method.
\end{abstract}

\noindent%
\vfill

\newpage
\spacingset{1.5} 
\section{Introduction}
\label{sec:intro}

Estimation of covariance and correlation matrices is of interest in many applications such as genetics \citep{butte2000discovering, wang2024estimation}, finance \citep{rothman2009generalized}, and neuroscience \citep{kim2023positive} as these estimates play crucial roles by capturing the interdependencies between variables.
However, 
when the number of variables exceeds the number of observations, traditional sample covariance matrices are inaccurate \citep{johnstone2001distribution} and, as an alternative, 
estimation of sparse covariance matrices where many elements are assumed to be zero has gained significant attention.
Despite substantial development in estimating sparse covariance matrices, existing methods still face several challenges such as accounting for the correlation among the elements of the sample covariance matrix, ensuring positive definiteness of the resulting estimator and unbiased estimation of the diagonal elements. These challenges compromise estimation accuracy and disqualify the estimator as a valid covariance matrix because it may not be positive definite.

A popular method to estimate sparse covariance matrices is to threshold the sample covariance or correlation matrix \citep{bickel2008covariance, rothman2009generalized, cai2011adaptive, jiang2013covariance}.
While this approach is popular due to its simplicity, 
it does not guarantee a positive definite (PD) estimator, posing challenges for specific downstream tasks. For example, quadratic discriminant analysis relies on class-specific covariance matrices and requires their inversion, which is only feasible if the matrices are PD. The lack of PD makes thresholding estimators unsuitable for such applications. 

Several approximation methods have been proposed to find PD thresholding estimators \citep{xue2012positive, rothman2012positive, wen2016positive, wen2021fast, kim2023positive, fatima2024two, wang2026adaptive}. 
However, these methods do not account for correlation among the elements of the sample covariance matrix.
Elements in a sample covariance matrix are often correlated, for example, in the multivariate Gaussian distribution with non-diagonal covariance matrix. When thresholding the sample covariance matrix, such correlation is not taken into account.
This is problematic because, for example, one element in a sample covariance matrix can be above the threshold simply due to its correlation with other non-zero elements even if its true covariance value is zero.

Several likelihood-based methods have been studied in \citet{lam2009sparsistency, bien2011sparse, wang2014coordinate, xu2022proximal} to address this correlation 
under the Gaussian assumption while producing a positive definite estimator for the covariance matrix.
However, the Gaussian likelihood is not convex, resulting in non-trivial optimization and potential for multiple local solutions \citep{zwiernik2017maximum}. 

The Gaussian likelihood for the inverse covariance matrix, on the other hand, is convex and sparse estimators for the inverse covariance matrix have been proposed \citep{Friedman2008, cai2011constrained, liu2017tiger}. 
However, the sparsity in the inverse covariance matrix does not translate into the sparsity in the covariance matrix, hence is of limited use if covariance matrix is the object of main interest. In addition, 
these likelihood-based methods for covariance (and inverse covariance) matrix estimation do not fix diagonal elements to the sample variance, potentially inducing bias, as will be discussed more in Section \ref{sec:meth}. 
Fixing the diagonal elements is also preferred when the data are scaled to have variance equal to one because the covariance matrix for the scaled data is equal to the correlation matrix whose diagonal elements should be equal to one.
Several attempts have been made for fixing diagonals of the sparse covariance matrix estimators \citep{cui2016sparse, liu2014sparse, wang2024estimation} but, as in thresholding estimators, these methods do not consider the correlation among the elements of the sample covariance matrix.

Given the drawbacks of existing methods for sparse covariance matrix estimation, our goal is to improve upon these methods in terms of accuracy by accounting for correlation among sample covariance matrix elements, preserving the positive definiteness of the estimator, and avoiding bias in the estimation of diagonal elements. Computationally, we aim to rely on convex optimization framework to avoid local solutions. Toward this goal, we propose a new sparse covariance matrix estimator based on penalized regression formulation of the sparse linear covariance model. Our approach accounts for correlation structure among sample covariance elements, while the corresponding optimization problem incorporates constraints that ensure positive definiteness of the solution as well as unbiasedness of the diagonal element (making them fixed as the sample variance).  We develop an algorithm to solve this optimization problem via the Alternating Directions Method of Multipliers (ADMM). In theoretical analysis, we demonstrate that our estimator is consistent in high-dimensional settings. In numerical studies on simulated datasets, we demonstrate that our estimator is more accurate than existing alternatives. We illustrate the use of our estimator on two downstream tasks: classification via quadratic discriminant analysis of healthy individuals from those with Parkinson's disease based on voice recording data and hierarchical clustering of gene expression data based on correlation distance.

\section{Methods}
\label{sec:meth}

\subsection{Notations}

Given a vector $\bx=(x_1, \ldots, x_p)^\top$, we use $\Vert\bx\Vert_1$, $\Vert\bx\Vert_2$ and $\Vert\bx\Vert_{\infty}$ to denote the vector $\ell_1$, $\ell_2$ and $\ell_{\infty}$ norms, respectively.
We use $\lambda_{\min}(\cdot)$ and $\lambda_{\max}(\cdot)$ to denote the smallest and largest eigenvalues of a matrix, respectively.
For a matrix $\bX\in\mathbb{R}^{p_1\times p_2}$, 
we let $\Vert\bX\Vert_1 = \sum_{ij}|X_{ij}|$, $\Vert\bX\Vert_2=\sqrt{\lambda_{\max}(\bX^\top \bX)}$, $\Vert\bX\Vert_{1,\infty}=\max_{j} \sum_{i=1}^{p_1}|\bX_{ij}|$ and $\Vert\bX\Vert_{\max}=\max_{ij}|X_{ij}|$ denote the matrix element-wise $\ell_1$ norm, spectral norm, matrix $\ell_1$ norm and element-wise max norm, respectively.
For a symmetric matrix $\bA\in\mathbb{R}^{p\times p}$,
$\text{vech}(\bA)=(A_{11}, A_{21}, A_{22},\ldots,A_{p1},\ldots,A_{pp})^\top$ represents the vectorization of the lower triangular portion of $\bA$ and $\text{vec}(\bA)$ represents the concatenation of columns in $\bA$. 
We also define an index set $[d] = \{k(k+1)/2: k = 1,\ldots,p \}$ which identifies the positions in $\text{vech}(\bA)$ where the diagonal elements of $\bA$ appear. Similarly, we define $[o]= \{1,\ldots,p(p+1)/2\}\setminus[d]$ for off-diagonal elements of $\bA$.


\subsection{Linear covariance model for element-wise estimation}

Consider a random vector $\by=(Y_1, Y_2, \ldots, Y_p)^\top \sim \mathbb{N}(\bzero,\bSigma^\ast)$. 
We consider the problem of estimating each entry in the symmetric and positive definite $\bSigma^\ast = (\sigma_{jk}^\ast)$ based on $n$ samples of $\by$, with $\bY$ denoting the $n \times p$ matrix of the data. Let $\bS = (s_{jk})$ denote the sample covariance matrix $\bS = \bY^T \bY/n$.
Since $\bS$ is an unbiased estimator of $\bSigma^\ast$,
\be \label{eq:linear_cov_noise}
\bS = \bSigma^\ast + \bE
\ee
with the error matrix $\bE=(\epsilon_{jk})$ such that $\mathbb{E}(\bE)=\bzero$.

We rewrite the model \eqref{eq:linear_cov_noise} as a vectorized form
\be \label{eq:linear_covreg}
\bs= \bsigma^\ast + \bepsilon
\ee
where $\bs=\text{vech}(\bS)$, $\bsigma^\ast=\text{vech}(\bSigma^\ast)$ and $\bepsilon=\text{vech}(\bE)$. 
Since the parameters in the model \eqref{eq:linear_covreg} are elements of the positive semi-definite matrix $\bSigma^\ast$, $\bsigma^\ast$ is constrained to the space
\begin{equation} \label{eq:pd_space}
\bTheta = \{\bsigma: \bSigma(\bsigma) \text{ is positive semi-definite}\},
\end{equation}
where $\bSigma(\bsigma)$ is the covariance matrix $\bSigma$ with $\bsigma$ as its elements.
Hence, estimation of $\bSigma^\ast$ can be cast as a constrained estimation of $\bsigma^\ast$ in the model \eqref{eq:linear_covreg}.
Both the ordinary least squares estimator and the maximum likelihood estimator of $\bsigma^\ast$ are equal to $\text{vech}(\bS)$ and the corresponding covariance matrix estimator $\bS$ is always positive semi-definite.  

\subsection{Sparse linear covariance model}
Let $\bV$ be the positive definite covariance matrix of $\bepsilon$ in \eqref{eq:linear_covreg} and suppose $\bV$ is known. 
Then, by multiplying $\bV^{-\frac{1}{2}}$ on both sides of \eqref{eq:linear_covreg}, the model can be written as a linear model
\be \label{eq:linear_covreg_gls}
\bV^{-\frac{1}{2}} \bs = \bV^{-\frac{1}{2}} \bsigma^\ast + \bV^{-\frac{1}{2}} \bepsilon
\ee
where $\bV^{-\frac{1}{2}} \bs$ is the response, $\bV^{-\frac{1}{2}}$ is the design matrix and $\bV^{-\frac{1}{2}} \bepsilon$ is the uncorrelated error. 
Although the sample covariance $\bs$ is the least squares estimator for the model \eqref{eq:linear_covreg_gls} and is also known as the maximum likelihood estimator under the Gaussian assumption \citep{watson1964note, zwiernik2017maximum}, its poor performance when the number of variables $p$ is large 
has been extensively studied \citep{pourahmadi2013high}. One remedy to address this limitation is to obtain a sparse covariance matrix estimator under the assumption that the covariance matrix $\bSigma^\ast$ is sparse, that is, many of its elements are zero \citep{bickel2008covariance, rothman2009generalized, rothman2012positive, lam2009sparsistency, bien2011sparse}. 

For a $p \times p$ symmetric matrix $\bSigma$ and $\bsigma=\text{vech}(\bSigma)$,
let $\bsigma_{[d]} \in \mathbb{R}^{p}$ and $\bsigma_{[o]} \in \mathbb{R}^{p(p-1)/2}$ be disjoint sub-vectors of $\bsigma$ with diagonal and off-diagonal elements in $\bSigma$, respectively. 
Similarly, let $\bs_{[d]}$ and $\bs_{[o]}$ be sub-vectors of $\bs$ and $\bepsilon_{[d]}$ and $\bepsilon_{[o]}$ be sub-vectors of $\bepsilon$.
Given $\wh{\bV}^{-\frac{1}{2}}$ as an estimator of $\bV^{-\frac{1}{2}}$, we propose the \textit{sparse linear covariance model (Sparse LCM)} estimator $\bSigma(\wh{\bsigma})$ obtained by solving the penalized regression problem
\be \label{eq:obj_covreg}
\wh{\bsigma} = \argmin_{\bsigma} \frac{1}{2n}\|\wh{\bV}^{-\frac{1}{2}} \bs-\wh{\bV}^{-\frac{1}{2}} \bsigma\|_2^2 + \lambda P(\bsigma_{[o]}) \quad \text{s.t.} \quad \bsigma_{[d]} = \bs_{[d]},
\ee
where $P(\cdot)$ is the sparsity-inducing penalty function and $\lambda>0$ is the tuning parameter.
As $\bsigma$ is the vector of elements in $\bSigma$, the sparsity of $\bsigma_{[o]}$ induced by the penalty term $P(\cdot)$ in \eqref{eq:obj_covreg} directly translates into the sparsity of $\bSigma$. 
For the penalty function, we use the $\ell_1$-penalty $P(\bsigma_{[o]})=\|\bsigma_{[o]}\|_1$ in this paper. However, other penalties such as the smoothly clipped absolute deviation (SCAD) penalty \citep{Fan2001} or the minimax concave penalty (MCP) \citep{zhang2010nearly} can also be considered.

By having the constraint $\bsigma_{[d]} = \bs_{[d]}$ in \eqref{eq:obj_covreg}, we fix the diagonal elements of $\bSigma$ to that of the sample covariance matrix. 
Fixing diagonals is not only common in other sparse covariance matrix estimators \citep{bickel2008covariance, rothman2012positive} but also,
we observe that not fixing the diagonals may induce unnecessary bias to the estimation of the diagonal elements, for example, in penalized likelihood estimators \citep{bien2011sparse, xu2022proximal}, as will be discussed further in Section 2.4.
In addition, fixing the diagonal elements is preferred for the scaled data where the variance is fixed to one.

The solution $\wh{\bsigma}$ in~\eqref{eq:obj_covreg} gives the covariance matrix estimator $\bSigma(\wh{\bsigma})$, and in Theorem \ref{thm3} we show that it is positive definite with high probability as the sample size increases. However, in finite sample settings, $\bSigma(\wh{\bsigma})$ may not be positive definite, thus in practice, we solve, for $\epsilon >0$,
\be \label{eq:obj_covreg_L1}
\wh{\bsigma}^{PD} = \argmin_{\bSigma(\bsigma) \succeq \epsilon \bI } \frac{1}{2n}\|\wh{\bV}^{-\frac{1}{2}} \bs-\wh{\bV}^{-\frac{1}{2}} \bsigma\|_2^2 + \lambda P(\bsigma_{[o]}) \quad \text{s.t.} \quad \bsigma_{[d]} = \bs_{[d]},
\ee
where $\bSigma(\bsigma) \succeq \epsilon \bI$ represents the positive definite constraint on $\bSigma$. Section~\ref{sec:optim} presents the corresponding optimization algorithm for the $\ell_1$-penalty.

In summary, the proposed Sparse LCM estimator accounts for correlation among sample covariance matrix elements, avoids bias in estimating diagonal elements, and is guaranteed to be positive definite.


\subsection{Relationship to prior works}

This formulation \eqref{eq:obj_covreg} is flexible to encompass many existing estimators of sparse covariance matrices depending on $\wh{\bV}$ and the penalty function. For example, 
when $\wh{\bV}=\bI_{p(p+1)/2}$, $\wh{\bsigma}$ gives the generalized thresholding estimator of $\bSigma^\ast$ \citep{rothman2009generalized}, including the hard thresholding estimator \citep{bickel2008covariance} when $P(\bsigma_{[o]})=\|\bsigma_{[o]}\|_0$, the soft thresholding estimator when $P(\bsigma_{[o]})=\|\bsigma_{[o]}\|_1$ and the $\ell_q$ regularized estimator \citep{wang2026adaptive} when $P(\bsigma_{[o]})=\|\bsigma_{[o]}\|_q$ for $0<q<1$. 
This implies that thresholding estimators do not consider the correlation among the elements of $\bs=\text{vech}(\bS)$ because $\wh{\bV}$ is a diagonal matrix. 
However, it is known that $\bS$ 
obtained from the $n$ samples of Gaussian $\by$ follows the Wishart distribution 
\citep{johnson2002applied} with the covariance matrix of $\text{vech}(\bS)$ as (page 317 of \citet{abadir2005matrix})
\be \label{eq:errorcov_gaussian}
{\rm Cov}\{\text{vech}(\bS)\} = \frac{2}{n} \bD_p^+ (\bSigma^\ast \otimes \bSigma^\ast) \bD_p^{+\top},
\ee
where $\bD_p^+=(\bD_p^\top \bD_p)^{-1} \bD_p^\top$ is the Moore-Penrose inverse of $\bD_p$ and $\bD_p$ is a duplication matrix (page 299 of \citet{abadir2005matrix}) which transforms $\text{vech}(\bA)$ into $\text{vec}(\bA)$ for a $p \times p$ symmetric matrix $\bA$ by $\bD_p \text{vech}(\bA) = \text{vec}(\bA)$ with $\text{vec}(\cdot)$ as the vectorization operator.
Details on this relationship between \eqref{eq:obj_covreg} and other thresholding estimators are discussed in Supplementary Materials \ref{rel_lcm_thres}.

The penalized regression \eqref{eq:obj_covreg} without the diagonal constraint also encompasses penalized likelihood estimator such as \citet{lam2009sparsistency, bien2011sparse}. Assuming that $\bV$ is known, we can show that the solution to the optimization \eqref{eq:obj_covreg} without the constraint $\bsigma_{[d]} = \bs_{[d]}$ is equivalent to the $\ell_1$-penalized likelihood estimator, which minimizes the negative Gaussian log-likelihood with $\ell_1$-penalty
\be \label{eq:log-lik}
\bSigma(\wh{\bsigma}) = \argmin_{\bSigma} \log \det (\bSigma) + \text{trace}(\bSigma^{-1} \bS) + \lambda \|\bP \circ \bSigma\|_1
\ee
where $\bP$ is a $p \times p$ matrix with all diagonal elements equal to zero and all off-diagonal elements equal to one and $\circ$ represents element-wise multiplication.
Detailed derivation of the relationship between \eqref{eq:obj_covreg} and the penalized likelihood estimator is discussed in Supplementary Materials \ref{rel_lcm_penal}.

\begin{figure}[hbt!]
	\centering
	\vspace*{6mm}
	\scalebox{0.6}{
	\mbox{
		\subfigure{
			\begin{overpic}[width=4.9in,angle=0]
				{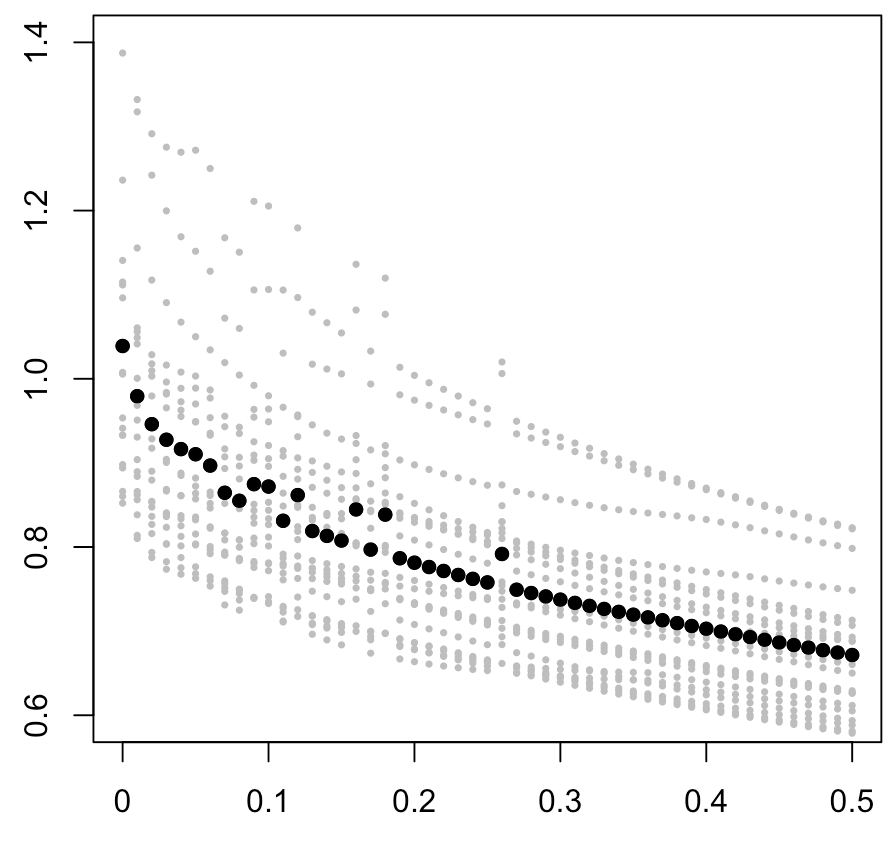}
            \put(48,98){\uline{\Large $n=100$}}
            \put(54,-3){\Large $\lambda$}
			\end{overpic}
		}
		\subfigure{
			\begin{overpic}[width=4.9in,angle=0]
				{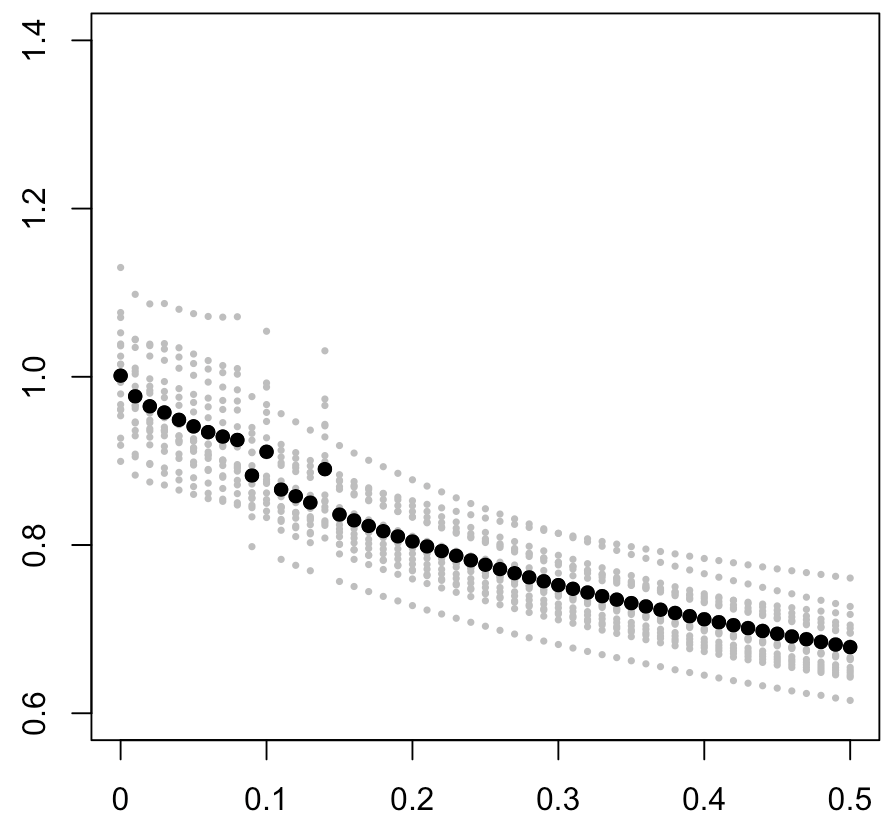}
            \put(46,98){\uline{\Large $n=1000$}}
            \put(54,-3){\Large $\lambda$}
			\end{overpic}
		}
	}
	
	}
	

\caption{Estimators of diagonal elements in $\bSigma^\ast$ for one dataset with $p=20$ and the sample size of $n=100$ or $n=1000$, simulated from $\mathbf{N}(\bzero, \bSigma^\ast)$. The covariance model for $\bSigma^\ast$ is the first-order moving average with the diagonal elements equal to one and the first off-diagonal elements equal to 0.5. For different values of penalty parameter $\lambda$ from 0 to 0.5, $\bSigma^\ast$ has been estimated by the $\ell_1$-penalized likelihood \eqref{eq:log-lik}. Estimators of 20 diagonal elements are shown in gray dots and the average of those 20 elements is shown in a black dot for each value of $\lambda$. The true parameter is equal to one and the bias in the diagonal estimators increases as $\lambda$ increases.}
\label{fig:bias_diag}
\end{figure}

However, existing penalized likelihood methods such as \citet{bien2011sparse} solve a non-convex problem which might suffer multiple local solutions. Also, the diagonal elements in those estimators are not fixed, which may induce bias to the diagonal elements, even if the penalty is imposed only on the off-diagonal elements.
To see this, we generated one dataset of size $n=100$ or $n=1000$ for 20 response variables $(p=20)$ from $\mathbf{N}(\bzero, \bSigma^\ast)$ where the diagonal elements of $\bSigma^\ast$ are all equal to one, the first off-diagonal elements are all equal to 0.5 and all other elements are zero. We estimated $\bSigma^\ast$ using the $\ell_1$-penalized likelihood \eqref{eq:log-lik}.
Estimators of the diagonal elements are plotted in Figure \ref{fig:bias_diag} for different values of the penalty parameter $\lambda$. Although the true parameter is equal to one, it is seen the bias in the diagonal elements increases as $\lambda$ increases.

\subsection{Optimization for positive definite covariance matrix}\label{sec:optim}


In constructing the linear model \eqref{eq:linear_covreg_gls}, $\bV^{-\frac{1}{2}}$ needs to be estimated. 
We propose to find $\wh{\bOmega}$, a consistent estimator of $\bOmega^\ast={\bSigma^\ast}^{-1}$, plug it in the inverse of \eqref{eq:errorcov_gaussian} as
\be \label{eq:errorinvcov_gaussian}
\wh{\bV}^{-1}= \frac{n}{2}\bD_p^\top (\wh{\bOmega} \otimes \wh{\bOmega}) \bD_p,
\ee
and then obtain $\wh{\bV}^{-\frac{1}{2}}$ as the square root matrix of $\wh{\bV}^{-1}$. 
Many consistent estimators of $\bOmega^\ast$ have been proposed such as the graphical LASSO \citep{Friedman2008} and the constrained $\ell_1$-minimization for inverse matrix estimation (CLIME) \citep{cai2011constrained} under different conditions. The choice of $\wh{\bOmega}$ and its theoretical justification will further be discussed in Section \ref{sec:theory}.

Next, we discuss the optimization of \eqref{eq:obj_covreg_L1} given $\wh{\bV}^{-\frac{1}{2}}$. Setting $P(\bsigma_{[o]})=\|\bsigma_{[o]}\|_1$, the problem can be rewritten as
\be \label{eq:obj_covreg_L1_2}
\wh{\bsigma}^{PD} = \argmin_{\bSigma(\bsigma) \succeq \epsilon \bI } \frac{1}{2n}\|\wh{\bV}^{-\frac{1}{2}} \bs-\wh{\bV}^{-\frac{1}{2}} \bsigma\|_2^2 + \lambda \|\bsigma_{[o]}\|_1 + \lambda_\infty \|\bsigma_{[d]} - \bs_{[d]}\|_1
\ee
where $\bSigma(\bsigma) \succeq \epsilon \bI$ represents the positive definite constraint on $\bSigma$ and $\lambda_\infty$ is large enough to shrink $\bsigma_{[d]}$ to be exactly equal to $\bs_{[d]}$.
Although the objective function corresponds to the LASSO problem for a linear model, existing algorithms for the LASSO, such as the coordinate descent, cannot account for the positive definite constraint.

Given that the optimization \eqref{eq:obj_covreg_L1_2} consists of three sub-problems: the least squares, the $\ell_1$-norm penalties and the positive definite constraint, we consider a distributed optimization scheme. 
In particular, we solve the problem using the consensus ADMM \citep{boyd2011distributed}, which reduces the formulation to a two-block ADMM scheme.
This two-block ADMM enjoys global convergence guarantees under convexity \citep{eckstein1992douglas} as well as computational efficiency from parallel solving of those sub-problems. 
Specifically, introducing $\btheta_1, \btheta_2, \btheta_3$ such that $\bsigma=\btheta_1=\btheta_2=\btheta_3$, 
the optimization \eqref{eq:obj_covreg_L1_2} can be reparametrized as
\bse
\wh{\bsigma}^{PD} = \argmin_{\bSigma(\btheta_2) \succeq \epsilon \bI } \frac{1}{2n}\|\wh{\bV}^{-\frac{1}{2}} \bs-\wh{\bV}^{-\frac{1}{2}} \btheta_1\|_2^2 + \lambda \|{\btheta_3}_{[o]}\|_1 + \lambda_\infty \|{\btheta_3}_{[d]} - \bs_{[d]}\|_1.
\ese
This problem is equivalent to solving the following multiple augmented Lagrangian problems:
\begin{align*}
&\argmin_{\btheta_1} \frac{1}{2n}\|\wh{\bV}^{-\frac{1}{2}} \bs-\wh{\bV}^{-\frac{1}{2}} \btheta_1\|_2^2 + \bnu_1^\top (\btheta_1-\bsigma) + \frac{\|\btheta_1-\bsigma\|_2^2}{2 \gamma},\\
&\argmin_{\bSigma(\btheta_2) \succeq \epsilon \bI } \bnu_2^\top (\btheta_2-\bsigma) + \frac{\|\btheta_2-\bsigma\|_2^2}{2 \gamma},\\
&\argmin_{\btheta_3} \lambda \|{\btheta_3}_{[o]}\|_1 + \lambda_\infty \|{\btheta_3}_{[d]} - \bs_{[d]}\|_1 + \bnu_3^\top (\btheta_3-\bsigma) + \frac{\|\btheta_3-\bsigma\|_2^2}{2 \gamma}
\end{align*}
where $\gamma$ determines the convergence rate and $\bnu_1,\bnu_2,\bnu_3$ are dual variables. 
Differentiating the first objective function with respect to $\btheta_1$ and setting it to zero gives an update rule for $\btheta_1$ as
\bse
\wh{\btheta}_1 = \bigg( \frac{1}{n}\wh{\bV}^{-1} + \frac{1}{\gamma}\bI \bigg)^{-1} \bigg( \frac{1}{n}\wh{\bV}^{-1} \bs + \frac{1}{\gamma}\bsigma - \bnu_1 \bigg).
\ese
For the second problem, we find a solution that meets the positive definite constraint by the projection onto the convex cone $\{\bSigma \succeq \epsilon \bI\}$ for a small $\epsilon>0$.
Let $\widecheck{\btheta}_2$ be the solution to the unconstrained problem and $\bSigma(\widecheck{\btheta}_2)=\sum_{j=1}^{p} \lambda_j \bqu_j \bqu_j^\top$ 
be the eigen-decomposition of $\bSigma(\widecheck{\btheta}_2)$. Then, the solution to the constrained problem is determined by $\wh{\btheta}_2$ that equates 
\bse
\bSigma(\wh{\btheta}_2) = \sum_{j=1}^{p} \max(\lambda_j, \epsilon) \bqu_j \bqu_j^\top.
\ese
Lastly, for the third problem, the subgradient equations for ${\btheta_3}_{[d]}$ and ${\btheta_3}_{[o]}$ are
\begin{align*}
\lambda_\infty \bu_{[d]} + {\bnu_3}_{[d]} + (1/\gamma) ({\btheta_3}_{[d]}-\bsigma_{[d]}) &=\bzero, \text{ and}\\ 
\lambda \bu_{[o]} + {\bnu_3}_{[o]} + (1/\gamma) ({\btheta_3}_{[o]}-\bsigma_{[o]}) &=\bzero
\end{align*}
where $\bu_{[d]} \in \mathbb{R}^p$ and $\bu_{[o]} \in \mathbb{R}^{p(p-1)/2}$ are the subgradient of $\|{\btheta_3}_{[d]} - \bs_{[d]}\|_1$ and $\|{\btheta_3}_{[o]}\|_1$, respectively.
Solving these equations gives $\wh{\btheta_3}_{[d]} = \bs_{[d]}$ for a sufficiently large $\lambda_\infty$ and $\wh{{\btheta_3}}_{[o]} = S_{\lambda \gamma}(\bsigma_{[o]} - \gamma \bnu_{[o]})$ where $S_{\lambda}(\ba)$ is the element-wise soft-thresholding operator at $\lambda$, that is, $\{S_{\lambda}(\ba)\}_j = \text{sign}(a_j)\times\max(|a_j|-\lambda, 0)$. 

By setting $\bleta_i=\gamma \bnu_i, (i=1,2,3)$, the scaled form of the above optimizations yields the consensus ADMM steps:
\begin{align*}
\btheta_1^{(k+1)}&=\argmin_{\btheta_1} \frac{1}{2n}\|\wh{\bV}^{-\frac{1}{2}} \bs-\wh{\bV}^{-\frac{1}{2}} \btheta_1\|_2^2 + \frac{1}{2\gamma}\|\btheta_1-\bsigma^{(k)}+\bleta_1^{(k)}\|_2^2,\\
\btheta_2^{(k+1)}&=\argmin_{\bSigma(\btheta_2) \succeq \epsilon \bI } \frac{1}{2\gamma}\|\btheta_2-\bsigma^{(k)}+\bleta_2^{(k)}\|_2^2,\\
\btheta_3^{(k+1)}&=\argmin_{\btheta_3} \lambda \|{\btheta_3}_{[o]}\|_1 + \lambda_\infty \|{\btheta_3}_{[d]} - \bs_{[d]}\|_1 + \frac{1}{2\gamma}\|\btheta_3-\bsigma^{(k)}+\bleta_3^{(k)}\|_2^2,\\
\bsigma^{(k+1)}&=(\btheta_1^{(k+1)}+\bleta_1^{(k)}+\btheta_2^{(k+1)}+\bleta_2^{(k)}+\btheta_3^{(k+1)}+\bleta_3^{(k)})/3,\\
\bleta_i^{(k+1)}&=\bleta_i^{(k)}+\btheta_i^{(k+1)}-\bsigma^{(k+1)}.
\end{align*}
The details of the proposed algorithm are given in Algorithm \ref{alg1}. 

\begin{algorithm}
	\caption{Positive definite approximation for the sparse linear covariance model}\label{alg1}
	\vspace{0.2 cm}
	Given $\wh{\bV} \in \mathbb{R}^{p(p+1)/2 \times p(p+1)/2}, \bS \in \mathbb{R}^{p \times p}$, 
    $\bsigma^{(0)}, \btheta_1^{(0)}, \btheta_2^{(0)}, \btheta_3^{(0)}, \bleta_1^{(0)}, \bleta_2^{(0)}, \bleta_3^{(0)}$, $\epsilon>0$ and $\gamma >0$, set $k=0$ and iterate until convergence.
	\begin{enumerate}
    \item $\btheta_1$ update: $\btheta_1^{(k+1)}=\bigg( \frac{1}{n}\wh{\bV}^{-1} + \frac{1}{\gamma} \bI_{p(p+1)/2} \bigg)^{-1} \bigg\{ \frac{1}{n}\wh{\bV}^{-1} \text{vech}(\bS) +\frac{1}{\gamma} (\bsigma^{(k)}-\bleta_1^{(k)}) \bigg\}$
    \item $\btheta_2$ update: set $\widecheck{\btheta}_2=\bsigma^{(k)}-\bleta_2^{(k)}$.
    With the eigen-decomposition of $\bSigma(\widecheck{\btheta}_2)=\sum_{j=1}^{p} \lambda_j \bqu_j \bqu_j^\top$, obtain $\btheta_2^{(k+1)}$ that equates
    \bse
    \bSigma(\btheta_2^{(k+1)}) = \sum_{j=1}^{p} \max(\lambda_j, \epsilon) \bqu_j \bqu_j^\top.
    \ese
    \item $\btheta_3$ update: fix ${{\btheta_3}_{[d]}}^{(i+1)}=\bs_{[d]}$ 
    and update ${{\btheta_3}_{[o]}}^{(k+1)}$ by
    \bse
    {{\btheta_3}_{[o]}}^{(k+1)} = S_{\lambda \gamma} ({\bsigma_{[o]}}^{(k)} - {{\bleta_3}_{[o]}}^{(k)})
    \ese
    where ${\bleta_3}_{[o]}=\gamma {\bnu_3}_{[o]}$ and $S_{\lambda}(\cdot)$ is the element-wise soft-thresholding operator at $\lambda$.
    \item $\bsigma$ update: $\bsigma^{(k+1)}=(\btheta_1^{(k+1)}+\bleta_1^{(k)}+\btheta_2^{(k+1)}+\bleta_2^{(k)}+\btheta_3^{(k+1)}+\bleta_3^{(k)})/3$
    \item $\bleta_1, \bleta_2, \bleta_3$ update: $\bleta_i^{(k+1)} = \bleta_i^{(k)} + (\btheta_i^{(k+1)}-\bsigma^{(k+1)})$ for $i=1,2,3$.
	\end{enumerate}
\end{algorithm}

\begin{Rem}
Algorithm \ref{alg1} provides an alternative for finding positive definite thresholding estimators, for example, similar to the positive definite soft thresholding estimator by \citet{xue2012positive}. However, unlike Algorithm \ref{alg1}, the diagonal elements in $\bSigma$ are not fixed in those existing methods. 
\end{Rem}
\begin{Rem} \label{rem2}
When $\wh{\bOmega}$ is available, an alternative approach to estimate $\bSigma$ is to use the heuristic ``dual likelihood" \citep{kauermann1996dualization}, which is motivated by interchanging the role of the covariance matrix and the inverse covariance matrix in the likelihood function. The penalized version of the dual likelihood method minimizes
\bse
-\log \det (\bSigma) + \text{trace}(\wh{\bOmega} \bSigma) + \lambda \|\bP \circ \bSigma\|_1.
\ese
Although this approach ensures the estimator of $\bSigma$ to be positive definite, it does not give a solution to the likelihood function. We compare the performance of this heuristic method with our proposed method in Section \ref{sec:simul}.
\end{Rem}

\subsection{Tuning parameter selection}

Tuning parameter selection for sparse covariance matrix estimation has relied broadly on either cross-validation or information criteria. On one hand, the sub-sampling method for minimizing the Frobenius risk by \citet{bickel2008regularized} is similar to cross-validation and has been popular for diverse sparse covariance estimators such as thresholding estimators \citep{bickel2008covariance, rothman2009generalized, cai2011adaptive, wen2021fast, wang2024estimation} and banding estimators \citep{bickel2008regularized, bien2016convex, bien2019graph}.
\citet{bien2011sparse} also used cross-validation that maximizes the likelihood function for the penalized likelihood estimator of sparse covariance matrices.
On the other hand, \citet{li2016sure} proposed Stein's unbiased risk estimation information criteria for bandable covariance matrices. \citet{kim2023positive} discussed that typical information criteria such as AIC and BIC can be used for the threshold selection of a thresholding estimator which is based on maximum likelihood estimation for Gaussian distribution.

For the sparse LCM estimator, we may encounter more tuning parameters in addition to $\lambda$ in \eqref{eq:obj_covreg_L1} because the estimation of $\bOmega^\ast$ 
may involve its own tuning parameter selection. 
For example, the graphical LASSO requires selection of the shrinkage parameter and similarly for the CLIME and others. 
Denoting the tuning parameter for $\wh{\bOmega}$ as $\rho$, we propose BIC to find $\lambda$ and $\rho$ which minimizes
\bse
n \cdot \log \det (\wh{\bSigma}_{\lambda, \rho}) + n \cdot \text{trace}(\bS \wh{\bSigma}_{\lambda, \rho}^{-1}) + \log(n) \cdot \sum_{j<k} \mathds{1}_{(\wh{\sigma}_{jk} \neq 0)}
\ese
where $\wh{\bSigma}_{\lambda, \rho}$ is the sparse LCM estimator obtained by solving \eqref{eq:obj_covreg_L1}, $\wh{\sigma}_{jk}$ is the $(j,k)$th element of $\wh{\bSigma}_{\lambda}$ and $\mathds{1}$ represents the indicator function. Similar information criteria have been discussed in \citet{danaher2014joint, maathuis2018handbook} for the Gaussian graphical model and in \citet{kim2023positive} for covariance matrix estimation.

\section{Theoretical Properties}
\label{sec:theory}
In this section, we derive a tight $\ell_2$ convergence rate for the off-diagonal elements of the proposed Sparse LCM solution to \eqref{eq:obj_covreg} in Theorem~\ref{thm1} and \ref{thm2}. In Theorem~\ref{thm3}, we demonstrate that it gives a positive definite estimator of $\bSigma^\ast$ with high probability as the sample size grows, thus asymptotically coinciding with $\widehat \bsigma^{\text{PD}}$ in~\eqref{eq:obj_covreg_L1}. The operator norm has previously been used to demonstrate consistency of thresholding estimators for the sparse covariance matrix \citep{bickel2008covariance, rothman2009generalized}. Consistency in terms of the Frobenius norm has also been considered for both thresholding estimator \citep{bickel2008covariance} and the penalized likelihood estimator \citep{lam2009sparsistency}. The convergence bound for Frobenius norm is not as tight as for the operator norm due to direct dependence on $p$, the dimension of $\by$, as a result of diagonal elements being not sparse. In the following, we thus focus on operator norm, with Frobenius norm bound following as Corollary~\ref{cor:frob}. 

To establish consistency of our estimator, we assume that the true covariance matrix $\bSigma^\ast$ is well conditioned as below.

\begin{Ass} \label{ass:eigen_bound}
There exist some constants $c$ and $C$ such that
\bse
0 < c \leq \lambda_{\min}(\bSigma^\ast) \leq \lambda_{\max}(\bSigma^\ast) \leq C < \infty.
\ese
\end{Ass}
Assumption \ref{ass:eigen_bound} implies that both the true covariance matrix and its inverse are positive definite.

First, we assume that the true $\bV^{-1}$ is known, that is, $\wh{\bV}^{-1}=\bV^{-1}$, and we derive the convergence rate of the solution to \eqref{eq:obj_covreg}.

\begin{proposition} \label{prop1}
Suppose Assumption \ref{ass:eigen_bound} holds. 
Let $\mathcal{S}$ be an index set $\{m:({\bsigma^\ast}_{[o]})_m \neq 0\}$ and let $s=|\mathcal{S}|$ denote the cardinality of $\mathcal{S}$. Let $\wt{\bsigma}$ be the minimizer of 
\bse
\frac{1}{2n}\|\bV^{-\frac{1}{2}} \bs-\bV^{-\frac{1}{2}} \bsigma\|_2^2 + \lambda \|\bsigma_{[o]}\|_1 \quad \text{s.t.} \quad \bsigma_{[d]} = \bs_{[d]}.
\ese
Define $\bV_o^{-\frac{1}{2}} \in \mathbb{R}^{p(p+1)/2 \times p(p-1)/2}$ as a sub-matrix of $\bV^{-\frac{1}{2}}$ containing all the $l$th columns of $\bV^{-\frac{1}{2}}$ for $l \in [o]$.
For $\lambda \geq 2 \|(\bV_o^{-\frac{1}{2}})^\top \bV_o^{-\frac{1}{2}} \bepsilon_{[o]}\|_\infty/n$, $\wt{\bsigma}_{[o]}$ satisfies
\bse
\|\wt{\bsigma}_{[o]} - {\bsigma^\ast}_{[o]}\|_2 \leq 3\lambda C^2 \sqrt{s}.
\ese
\end{proposition}


Proposition \ref{prop1} states that the $\ell_2$ convergence rate of $\wt{\bsigma}_{[o]}$ is bounded by a factor of $\lambda$ and also depends on $s$, the level of sparsity in off-diagonals of $\bSigma^\ast$ and this convergence rate is similar to the deterministic estimation error bound for the LASSO (e.g. Theorem 11.1 of \citet{hastie2015statistical}). 

Based on Proposition \ref{prop1}, we obtain a probabilistic bound for the $\ell_2$ convergence rate of $\wt{\bsigma}_{[o]}$ with true $\bV^{-1}$.

\begin{Th}\label{thm1}
Suppose assumptions in Proposition \ref{prop1} hold. 
Let $\by_i$ be the $i$th observation of $\by$ and $\xi_{il}$ be the $l$th element of $n^{-1}(\bV_o^{-\frac{1}{2}})^\top \bV_o^{-\frac{1}{2}} \text{vech}(\by_i \by_i^\top - \bSigma^\ast)_{[o]} \in \mathbb{R}^{p(p-1)/2}$.
Let $\| \cdot \|_{\psi_1}$ denote the sub-exponential norm and define $K_1=\max_{i,l} \| \xi_{il} \|_{\psi_1}$.
For some constants $c_1>0, C_1>\sqrt{2c_1^{-1}}$, set $\lambda$ as 
\bse
\lambda = 2 K_1 C_1 \sqrt{\frac{\log p}{n}}.
\ese
Then, with probability at least $1- p^{2-c_1 C_1^2}$,
\bse
\|\wt{\bsigma}_{[o]} - {\bsigma^\ast}_{[o]}\|_2 \leq 6 K_1 C_1 C^2 \sqrt{\frac{s \log p}{n}}.
\ese
\end{Th}

It is worth comparing the result of Theorem \ref{thm1} with the Frobenius norm consistency of other estimators such as the thresholding estimator \citep{bickel2008covariance} 
or the penalized likelihood estimator \citep{lam2009sparsistency}, which include $(p \log p/n)^{1/2}$ in the convergence rate. By bounding only the off-diagonal part in $\bSigma$, which is the source of sparsity, $\wt{\bsigma}_{[o]}$ does not suffer the inclusion of $p$ in the convergence rate. Sub-exponentiality of $\xi_{il}$ in Theorem \ref{thm1} and its uniform bound $K_1$ is discussed in Lemma \ref{lem1} in Supplementary Materials \ref{thm1_proof}.


Next, we consider the estimation error of the solution to \eqref{eq:obj_covreg} with $\wh{\bV}^{-1}$. 
Particularly, we consider $\wh{\bV}^{-1}$ constructed by plugging the CLIME estimator \citep{cai2011constrained} of $\bOmega^\ast$ in \eqref{eq:errorinvcov_gaussian}. For this, we adopt the same parameter space for the inverse covariance matrix as proposed by \citet{cai2011constrained} as below.

\begin{Ass} \label{ass:invcov_sparsity}
$\bOmega^\ast
\in \{\bOmega: \|\bOmega\|_{1,\infty}\leq M, \max_{1 \leq i \leq p} \sum_{j=1}^p \omega_{ij}^q \leq t \}$ for $0 \leq q < 1$.
\end{Ass}

Let $\wt{\bOmega}$ be the solution to the optimization
\be \label{eq:clime}
\min\|\bOmega\|_1 \quad \text{subject to} \quad \|\bS \bOmega - \bI\|_{\max} \leq \rho
\ee
and let $\wh{\bOmega}$ be a thresholding estimator whose $(j,k)$th element $\hat{\omega}_{jk}$ is determined by
\be \label{eq:clime_threshold}
\hat{\omega}_{jk} = \tilde{\omega}_{jk} \mathds{1}_{(\tilde{\omega}_{jk} \geq \tau_n)}
\ee
where 
$\tilde{\omega}_{jk}$ is the $(j,k)$th element of $\wt{\bOmega}$ and
$\tau_n \geq 4 C_0 M^2 \sqrt{\log p/n}$ with $C_0$ as defined in \citet{cai2011constrained}.
The model selection consistency of $\wh{\bOmega}$ was discussed in \citet{cai2011constrained} under the following condition.

\begin{Ass} \label{ass:beta_min}
$\omega_{\min}^\ast \geq 2 \tau_n$ where $\omega_{\min}^\ast$ is defined as
\bse
\omega_{\min}^\ast = \min_{(j,k)\in \mathcal{S}(\bOmega^\ast)}|\omega_{jk}^\ast|
\ese
with $\omega_{jk}^\ast$ being the $(j,k)$th element of $\bOmega^\ast$ and $\mathcal{S}(\bOmega^\ast) = \{(j,k):\omega_{jk}^\ast \neq 0\}$.
\end{Ass}

\begin{Th}\label{thm2}
Suppose 
assumptions in Theorem \ref{thm1} and Assumption \ref{ass:invcov_sparsity}-\ref{ass:beta_min} hold and
set $\wh{\bV}^{-1}= (n/2)\bD_p^\top (\wh{\bOmega} \otimes \wh{\bOmega}) \bD_p$ with $\wh{\bOmega}$ as defined by \eqref{eq:clime} and \eqref{eq:clime_threshold}.
Let $\wh{\bsigma}$ be the minimizer of \eqref{eq:obj_covreg}.
For $\lambda = o(\sqrt{\log p/n})$ and $\tau>0$,
if $s=o(\sqrt{n/\log p})$ and $t^2=o(\sqrt{n/\log p})$,
there exists a constant $C^\ast$ such that 
\bse
\|\wh{\bsigma}_{[o]} - {\bsigma^\ast}_{[o]}\|_2 \leq C^\ast \sqrt{\frac{s \log p}{n}}
\ese
with probability $1 - 8 p^{\min(2-c_1 C_1^2, -\tau)}$.
\end{Th}

Theorem \ref{thm2} shows that, when $\wh{\bV}^{-1}$ is constructed based on the CLIME estimator $\wh{\bOmega}$, the $\ell_2$ convergence rate of our proposed estimator $\wh{\bsigma}_{[o]}$ has a tight bound, similar to the bound in Theorem \ref{thm1} up to a constant. 

We next establish an operator norm bound on the convergence rate for $\bSigma(\wh{\bsigma})$, which accounts for diagonal elements.
\begin{Th}\label{thm3}
Suppose assumptions in Theorem \ref{thm2} hold. Then,
\bse
\|\bSigma(\wh{\bsigma})-\bSigma^\ast\|_2 = O_p \bigg( s \sqrt{\frac{\log p}{n}} \bigg).
\ese
\end{Th}
In addition to establishing operator norm consistency, Theorem \ref{thm3} implies that the solution to the optimization \eqref{eq:obj_covreg} satisfies the PD constraint \eqref{eq:pd_space} as $n$ increases. Hence, $\wh{\bsigma}$ from \eqref{eq:obj_covreg} is asymptotically equal to $\wh{\bsigma}^{PD}$ from \eqref{eq:obj_covreg_L1}.

Finally, we establish Frobenius norm consistency as a corollary since $\wh{\bsigma}_{[d]}=\bs_{[d]}$ and 
$\|\bs_{[d]} - {\bsigma^\ast}_{[d]}\|_2 = O_p(\sqrt{p/n})$. As mentioned above, the Frobenius norm convergence bound is less tight than the operator norm bound, consistent with prior theoretical analyses of covariance matrix estimators \citep{bickel2008covariance, lam2009sparsistency}.
\begin{Cor}\label{cor:frob}
Suppose assumptions in Theorem \ref{thm2} hold.
$\bSigma(\wh{\bsigma}_{[d]}, \wh{\bsigma}_{[o]})$, the estimator of $\bSigma^\ast$ constructed by $\wh{\bsigma}=(\wh{\bsigma})$, satisfies
\bse
\|\bSigma(\wh{\bsigma})-\bSigma^\ast\|_F = O_p \bigg(\sqrt{\frac{p + s\log p}{n}} \bigg). 
\ese
\end{Cor}

\section{Simulation}
\label{sec:simul}

\subsection{Simulation settings}

We generate 20 datasets, each with sample size $n \in \{50, 100\}$ and the number of variables $p \in \{50, 100\}$ from $\Normal_p(\bzero, \bSigma^\ast)$. 
Off-diagonal elements of $\bSigma^\ast$ are determined by one of the following models, which are also considered in \citet{bien2011sparse, xu2022proximal}.
\begin{itemize}
    \item Moving average, MA(1): $\sigma_{jk}^\ast=0.4$ if $|j-k|=1$ and $\sigma_{jk}^\ast=0$ if $|j-k|>1$. 
    \item Random: each off-diagonal element of $\bSigma^\ast$ is non-zero with probability 0.02. 
    \item Hub: $\sigma_{jk}^\ast$ is non-zero for $j$ such that $\text{mod}(j,p/5)=1$ and $k \in \{j+1,\ldots,j+p/5-1\}$ and $\sigma_{kj}^\ast=\sigma_{jk}^\ast$. Otherwise, $\sigma_{jk}^\ast = 0$ if $j \neq k$.
\end{itemize}
For the random model and the hub model, 
the non-zero elements are randomly determined to be either $1$ or $-1$ with equal probability. 
To ensure $\bSigma^\ast$ is positive definite, the diagonal elements of $\bSigma^\ast$ are set to a constant so that the resulting matrix has the ratio of its largest to smallest eigenvalue equal to $p$. 
This setup on the eigenvalues of $\bSigma^\ast$ is common in high dimensional covariance matrix estimation and has been adopted in simulation studies by \citet{rothman2008sparse, bien2011sparse, xu2022proximal}.
\subsection{Performance evaluation}

We evaluate the finite sample performance of our proposed estimator \eqref{eq:obj_covreg_L1} with other sparse covariance matrix estimators.
Specifically, we compare our proposed sparse linear covariance model estimator 
with $\wh{\bV}^{-1}$ obtained by plugging the CLIME estimator $\wh{\bOmega}$ in \eqref{eq:errorinvcov_gaussian}, referred to as \texttt{SpLCM}, with other four existing methods, including
($\romannumeral 1$) the positive definite soft thresholding estimator by \citet{rothman2012positive}, referred to as \texttt{Soft}; ($\romannumeral 2$) the sparse covariance matrix from the penalized likelihood estimation by \citet{bien2011sparse}, referred to as \texttt{SpCov}; ($\romannumeral 3$) the likelihood-based estimator from a proximal distance algorithm by \citet{xu2022proximal}, referred to as \texttt{ProxCov}; ($\romannumeral 4$) the penalized dual likelihood estimator discussed in Remark \ref{rem2}, referred to as \texttt{PenDual}.
\texttt{Soft} and \texttt{SpCov} are implemented using the R package \texttt{PDSCE} and \texttt{covglasso}, respectively, and \texttt{ProxCov} is implemented using the code provided by \citet{xu2022proximal}. 
When $p \geq n$, \texttt{SpCov} and \texttt{ProxCov} estimators are not available because these methods require the sample covariance matrix $\bS$ to be full-rank. In such cases, \citet{bien2011sparse} suggested to use $\bS + \epsilon \bI$ with some $\epsilon>0$ instead of $\bS$ and we follow their suggestion with $\epsilon=0.001$.
The tuning parameter for \texttt{Soft} and \texttt{ProxCov} is selected by cross validation.
For \texttt{SpCov} and \texttt{SpLCM}, we used BIC to select the tuning parameter.

\begin{figure}[hbt!]
	\centering
	
	\scalebox{0.4}{
	\mbox{
		\subfigure{
			\begin{overpic}[width=4.9in,angle=0]
				{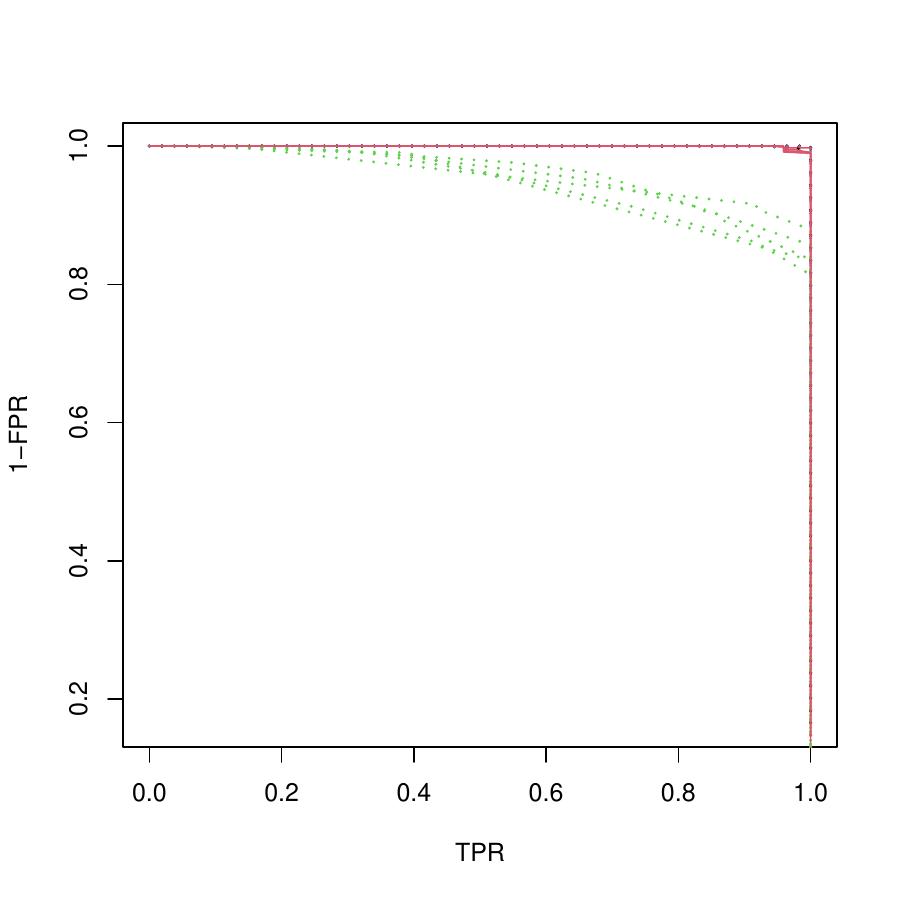}
				\put(-10,30){\rotatebox{90}{\LARGE{$n=100, p=50$}}}
				\put(46,94){\LARGE{\uline{MA(1)}}}
			\end{overpic}
		}
		\subfigure{
			\begin{overpic}[width=4.9in,angle=0]
				{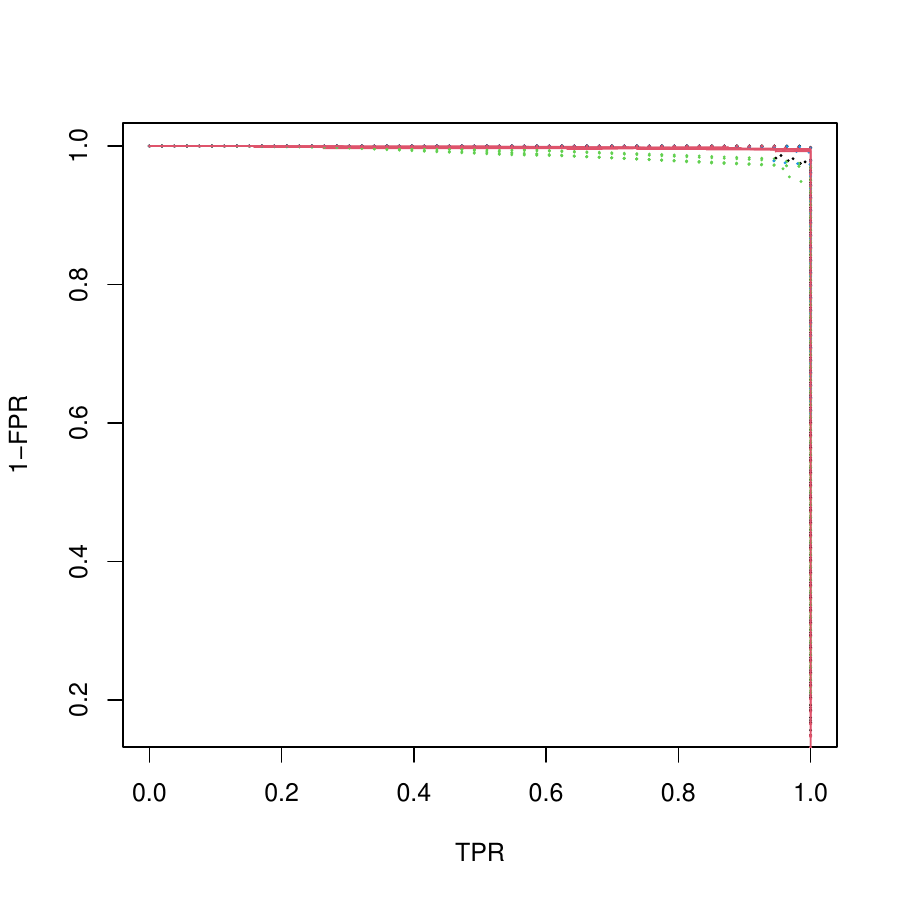}
				\put(42,94){\LARGE{\uline{Random}}}
			\end{overpic}
		}
        \subfigure{
			\begin{overpic}[width=4.9in,angle=0]
				{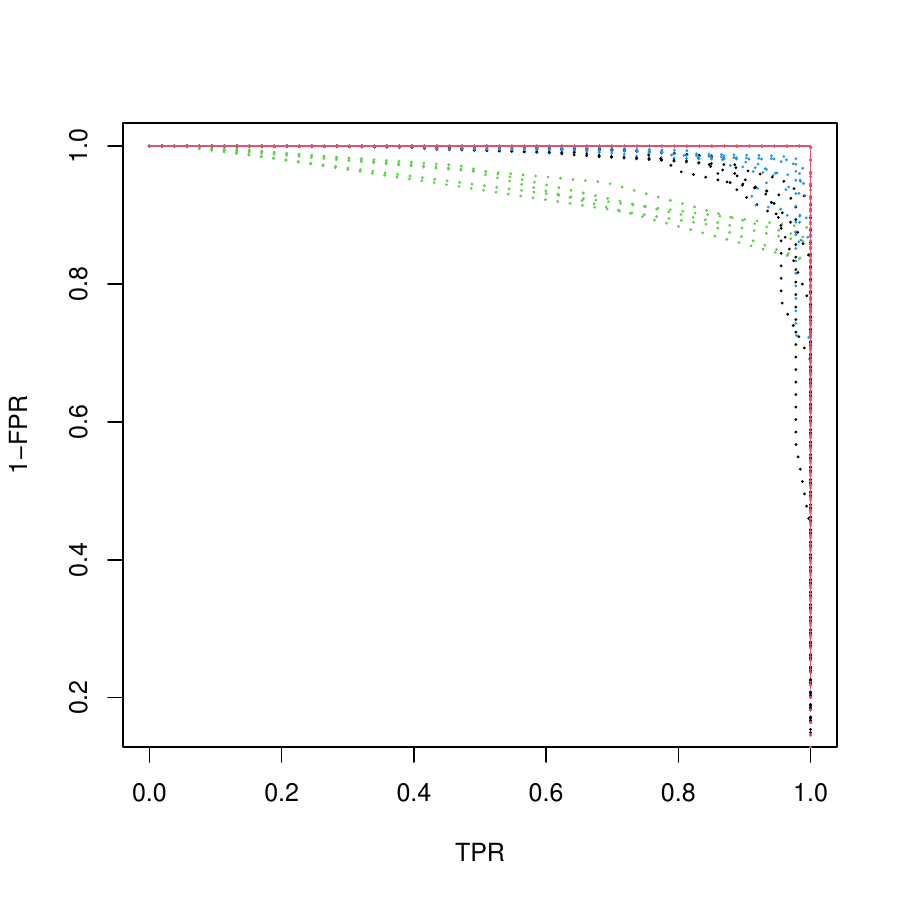}
				\put(46,94){\LARGE{\uline{Hub}}}
			\end{overpic}
		}
		
	}
	
	}
	
	
	\scalebox{0.4}{
	\mbox{
		\subfigure{
			\begin{overpic}[width=4.9in,angle=0]
				{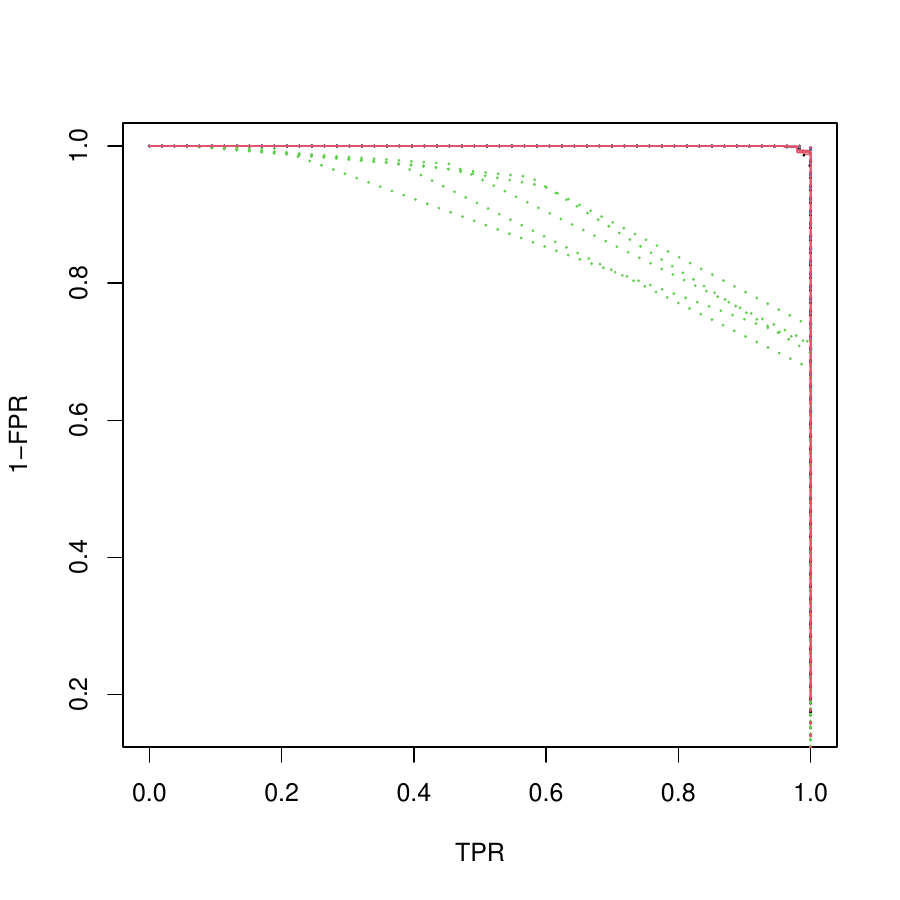}
				\put(-10,30){\rotatebox{90}{\LARGE{$n=100, p=100$}}}
			\end{overpic}
		}
		\subfigure{
			\begin{overpic}[width=4.9in,angle=0]
				{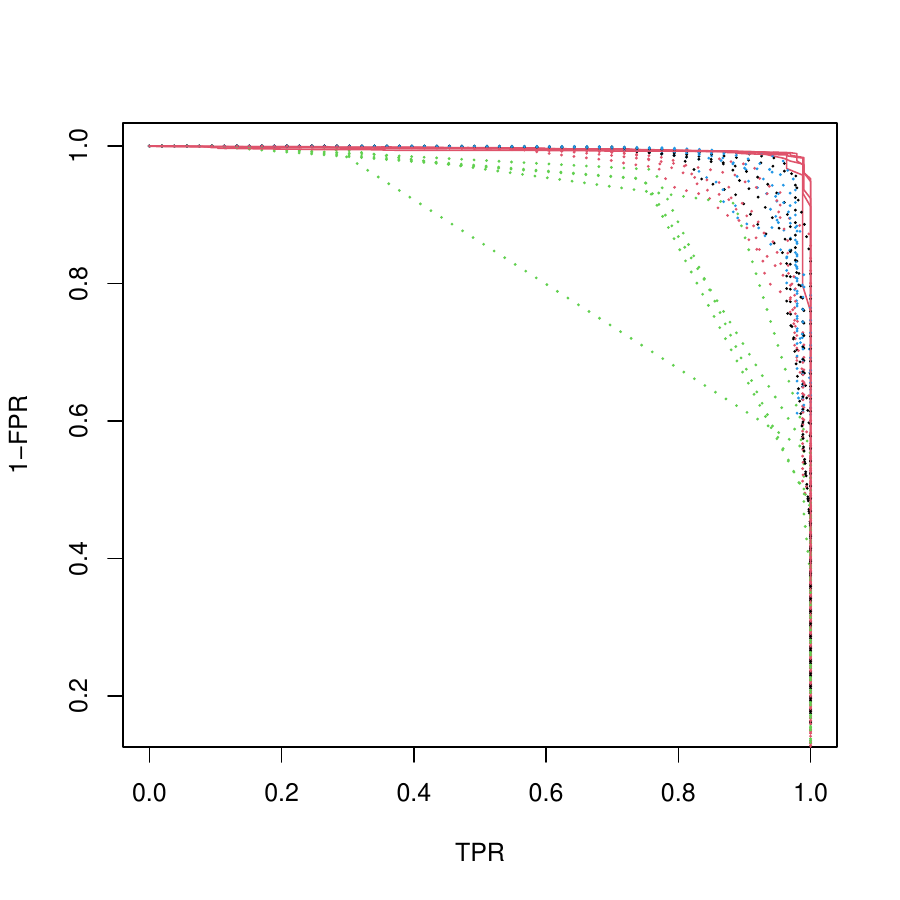}
			\end{overpic}
		}
        \subfigure{
			\begin{overpic}[width=4.9in,angle=0]
				{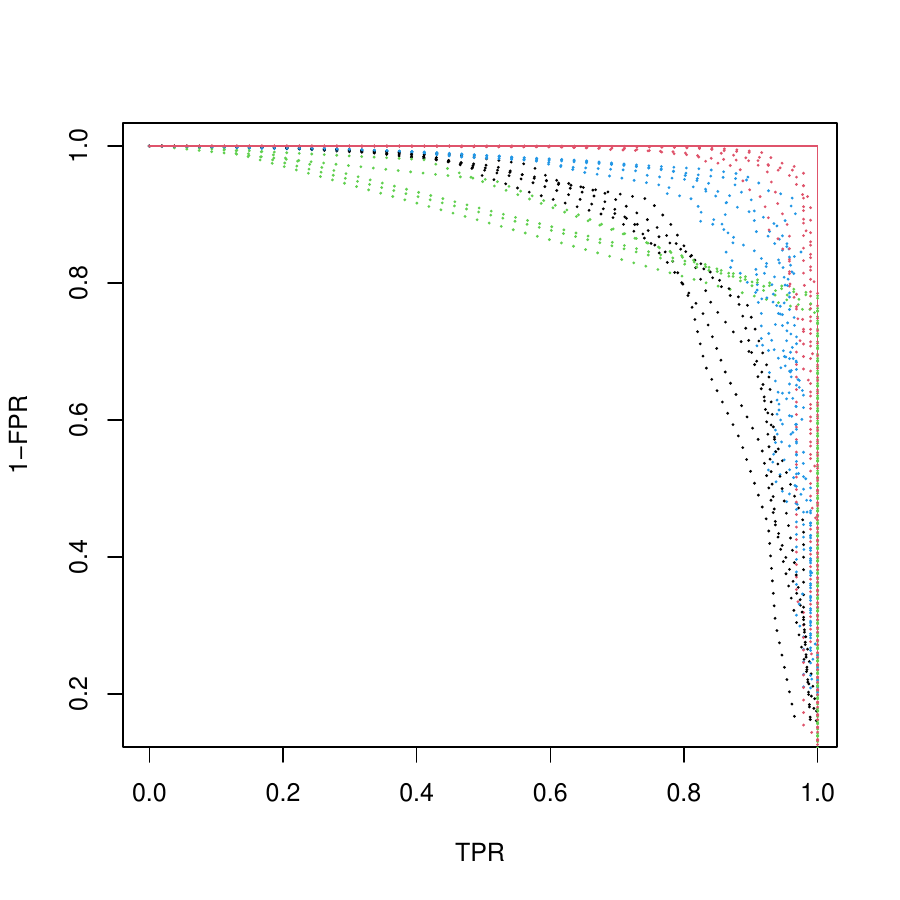}
			\end{overpic}
		}
		
	}
	
	}
	
	
	\scalebox{0.4}{
	\mbox{
		\subfigure{
			\begin{overpic}[width=4.9in,angle=0]
				{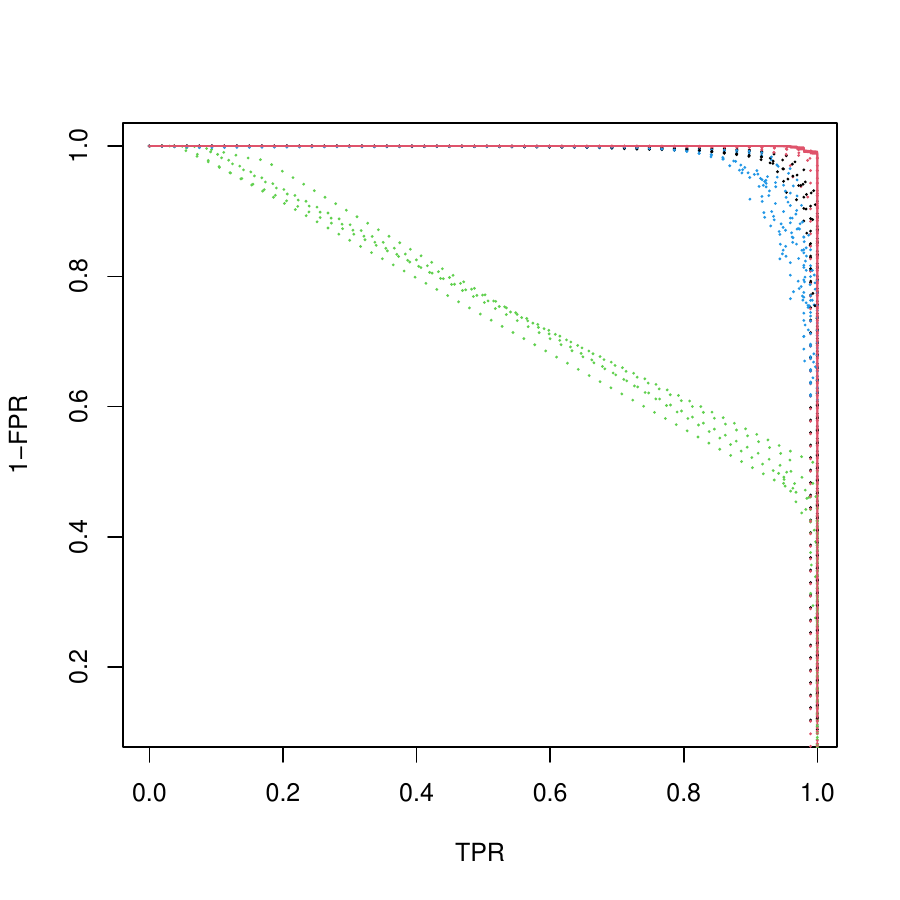}
				\put(-10,30){\rotatebox{90}{\LARGE{$n=50, p=100$}}}
			\end{overpic}
		}
		\subfigure{
			\begin{overpic}[width=4.9in,angle=0]
				{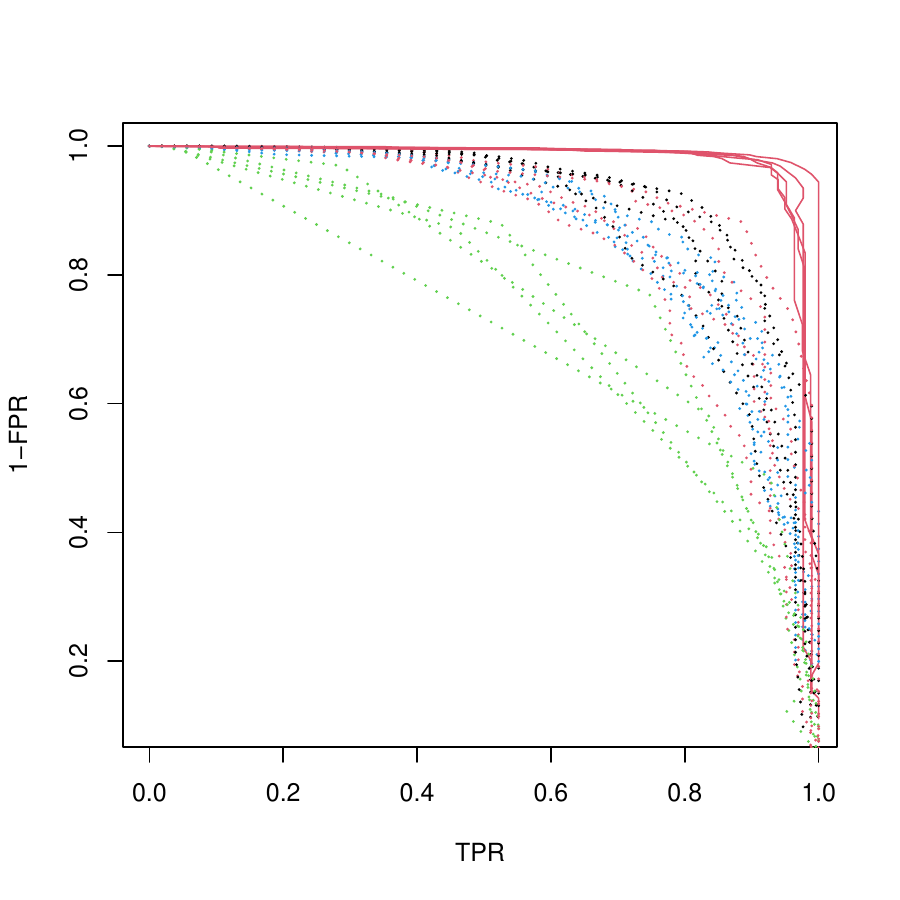}
			\end{overpic}
		}
        \subfigure{
			\begin{overpic}[width=4.9in,angle=0]
				{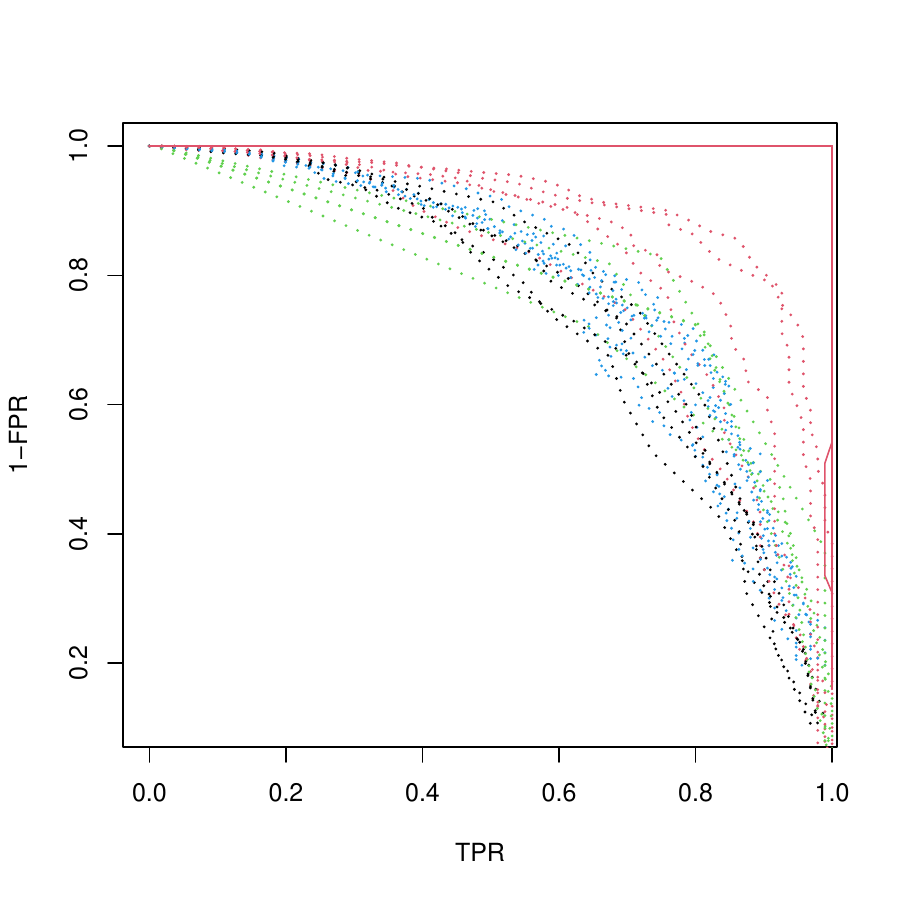}
			\end{overpic}
		}
		
	}
	
	}

\caption{ROC curves of \texttt{Soft} (black dotted), \texttt{SpCov} (green dotted), \texttt{ProxCov} (blue dotted), the proposed \texttt{SpLCM} (red dotted) and the oracle \texttt{SpLCM(O)} (red solid) for 5 simulated datasets. The values on x-axis represent sensitivity and the values on y-axis represent specificity.}
\label{fig:roc}
\end{figure}

We evaluate the performance in two aspects: estimation accuracy and support recovery. 
For estimation performance, denoting $\wh{\bSigma}$ as the estimator of $\bSigma^\ast$, we compare the $\ell_2$ norm distance of the off-diagonal elements $\|\text{vech}(\wh{\bSigma} - \bSigma^\ast)_{[o]}\|_2$. We also compare the estimation error of the covariance matrix estimator by measuring the Frobenius norm distance $\|\wh{\bSigma} - \bSigma^\ast\|_F$ and the operator norm distance $\|\wh{\bSigma} - \bSigma^\ast\|_2$. These distance measures were used in Section \ref{sec:theory} to show the convergence of our proposed estimator.
For support recovery performance, we compare the true positive rate and the false positive rate. 
These are standard measures for support recovery \citep{rothman2009generalized, xue2012positive} and defined as below:
\begin{align*}
    \text{True positive rate} & = \frac{\#\{(i,j): \hat{\sigma}_{ij} \neq 0,\; \sigma_{ij} \neq 0 \}}{\#\{(i,j): \sigma_{ij} \neq 0\}} \\
    \text{False positive rate} & = \frac{\#\{(i,j): \hat{\sigma}_{ij} \neq 0,\; \sigma_{ij} = 0 \}}{\#\{(i,j): \sigma_{ij} = 0\}}
\end{align*}
where $\#$ denotes the number of elements in a set. An estimator with higher true positive rate and lower false positive rate is preferred.
Estimation performance and support recovery performance of all competing methods are summarized in Table \ref{tb:total_cov} and Table \ref{tb:tprfpr}, respectively. In addition, we compare the sample covariance matrix (\texttt{Sample}) and the oracle sparse LCM estimator (\texttt{SpLCM(O)}) where the true $\bV^{-1}$ is used in the optimization \eqref{eq:obj_covreg_L1} instead of $\wh{\bV}^{-1}$. 
We also visualize the support recovery performance by plotting the receiver operating characteristic (ROC) curves for 5 simulated datasets in Figure \ref{fig:roc}.

First, we discuss the benefits of considering the correlation among the elements of $\bS$. 
Observe that the solution to our proposed optimization \eqref{eq:obj_covreg} encompasses the soft thresholding estimator when $\wh{\bV}^{-1}= (n/2)\bD_p^\top (\bI_p \otimes \bI_p) \bD_p$ 
which is a diagonal matrix and, hence, does not consider the correlation among the elements of $\bS$. 
Hence, by comparing the soft thresholding estimator \texttt{Soft} with the sparse LCM estimator \texttt{SpLCM} and \texttt{SpLCM(O)}, we can measure the effect of considering the correlation among the elements of $\bS$.
As seen in Table \ref{tb:total_cov}, \texttt{SpLCM(O)} exhibits much lower Frobenius norm distance than \texttt{Soft} in all simulation settings and in all covariance models considered. 
These results suggest that the estimation of sparse covariance matrices can be improved by considering the correlation among the elements of $\bS$.

Next, we compare the estimation performance of all covariance matrix estimators.
In Table \ref{tb:total_cov}, all sparse covariance matrix estimators consistently show lower Frobenius norm distance than the sample covariance matrix across all simulation settings.
Specifically, our proposed \texttt{SpLCM} exhibits the lowest (or comparable to the lowest) distance from the true covariance matrix across all simulation settings, indicating the highest accuracy in covariance matrix estimation.
When $n=50$ and $p=100$, for the random model and the hub model, the level of the noise in the data is excessive and very sparse matrices are selected by all methods except \texttt{ProxCov}, leading to similar estimators to each other. However, for the MA(1) model, \texttt{SpLCM} still outperforms others by showing the lowest distance from $\bSigma^\ast$.

In terms of the support recovery performance, \texttt{SpLCM} often shows relatively higher true positive rate and lower false positive rate than others in Table \ref{tb:tprfpr}, particularly for the MA(1) model and the hub model.
Its superior support recovery performance is also demonstrated by the ROC curves in Figure \ref{fig:roc}. In many cases, for example in the hub model, \texttt{SpLCM} shows relatively higher sensitivity and specificity than other methods across different values of tuning parameter.

Notably, although not available in practice, 
\texttt{SpLCM(O)} consistently shows the best performance in all aspects in Table \ref{tb:total_cov}, Table \ref{tb:tprfpr} and Figure \ref{fig:roc}, indicating the strength of the sparse LCM over the thresholding estimators or the likelihood-based estimators. Despite the superior performance of \texttt{SpLCM} over other covariance matrix estimators, the gap between \texttt{SpLCM} and \texttt{SpLCM(O)} suggests potential for future research to find a better estimator of $\bV^{-1}$ in the sparse LCM.
For example, the CLIME estimation procedure can be adopted to estimate $\bV^{-1}$ directly instead of estimating $\bOmega^\ast$. In this approach, controlling the computational complexity will be of major interest because the dimension of $\bV^{-1}$ is much larger than that of $\bOmega^\ast$ and we leave this for future research.

\begin{table}[!t]
\centering
\scalebox{0.8}{
\begin{tabular}{llrrrrrrrrr}
  \hline \hline
  & & \multicolumn{3}{c}{\uline{\quad \; $\|\text{vech}(\wh{\bSigma} - \bSigma^\ast)_{[o]}\|_2$ \; \quad}} & \multicolumn{3}{c}{\uline{\quad \quad \; $\|\wh{\bSigma} - \bSigma^\ast\|_F$ \;  \quad \quad}} & \multicolumn{3}{c}{\uline{\quad \quad \; $\|\wh{\bSigma} - \bSigma^\ast\|_2$ \;  \quad \quad}} \\[1.0ex]
  & method & MA(1) & Random & Hub & MA(1) & Random & Hub & MA(1) & Random & Hub \\[1.0ex] \hline
  $n=100$ & \texttt{Sample} & 2.93 & 8.30 & 11.00 & 4.23 & 11.95 & 15.87 & 1.70 & 4.52 & 5.94   \\ [1.0ex]
  $p=50$ & \texttt{Soft} & 1.25 & 2.77 & 4.64 & 1.95 & 4.53 & 7.28 & 0.60 & 1.62 & 2.68   \\ [1.0ex]
  & \texttt{SpCov} & 1.32 & 2.78 & 2.64 & 2.82 & 5.14 & 6.28 & 0.81 & 1.87 & 2.42 \\ [1.0ex]
  & \texttt{ProxCov} & 0.78 & \textbf{2.09} & 3.41 & 1.61 & 4.50 & 6.48 & 0.57 & 1.67 & 2.47 \\ [1.0ex]
  & \texttt{PenDual} & 1.45 & 3.31 & 6.56 & 2.48 & 5.83 & 11.69 & 0.72 & 2.47 & 3.74 \\ [1.0ex]
  & \texttt{SpLCM} & \textbf{0.77} & 2.29 & \textbf{2.07} & \textbf{1.38} & \textbf{3.99} & \textbf{4.31} & \textbf{0.50} & \textbf{1.37} & \textbf{1.69} \\ [1.0ex]
  \hdashline
  & \texttt{SpLCM(O)} & 0.62 & 1.83 & 1.20 & 1.22 & 3.47 & 3.58 & 0.46 & 1.21 & 1.41 \\ [1.0ex]
  \hline
  $n=100$ & \texttt{Sample} & 5.80 & 22.09 & 31.49 & 8.29 & 31.55 & 44.99 & 2.68 & 9.42 & 13.44   \\ [1.0ex]
  $p=100$ & \texttt{Soft} & 1.90 & \textbf{7.04} & 8.96 & 2.94 & \textbf{10.89} & 14.19 & 0.65 & 2.62 & 4.62   \\ [1.0ex]
  & \texttt{SpCov} & 3.39 & 8.73 & 9.56 & 5.25 & 13.64 & 15.03 & 0.92 & 3.21 & 4.97 \\ [1.0ex]
  & \texttt{ProxCov} & 2.22 & 8.68 & 11.10 & 5.01 & 21.58 & 25.81 & 1.02 & 4.43 & 6.89 \\ [1.0ex]
  & \texttt{PenDual} & 2.25 & 8.63 & 9.75 & 3.74 & 13.73 & 19.75 & 0.78 & 3.45 & 5.86 \\ [1.0ex]
  & \texttt{SpLCM} & \textbf{1.12} & 7.28 & \textbf{7.05} & \textbf{1.98} & 11.22 & \textbf{11.87} & \textbf{0.53} & \textbf{2.54} & \textbf{3.79} \\ [1.0ex]
  \hdashline
  & \texttt{SpLCM(O)} & 0.88 & 4.70 & 1.81 & 1.71 & 8.00 & 6.91 & 0.51 & 1.95 & 2.07 \\ [1.0ex]
  \hline
  $n=50$ & \texttt{Sample} & 8.29 & 31.37 & 44.73 & 11.84 & 44.80 & 63.91 & 4.15 & 14.90 & 21.05   \\ [1.0ex]
  $p=100$ & \texttt{Soft} & 2.67 & \textbf{9.21} & \textbf{9.69} & 4.12 & \textbf{14.43} & \textbf{16.35} & 0.85 & \textbf{3.33} & \textbf{5.23}   \\ [1.0ex]
  & \texttt{SpCov} & 3.91 & 9.80 & 9.75 & 5.83 & 15.29 & 16.59 & 1.02 & 3.48 & 5.31 \\ [1.0ex]
  & \texttt{ProxCov} & 3.79 & 11.27 & 12.39 & 8.34 & 30.18 & 37.72 & 1.50 & 5.78 & 8.09 \\ [1.0ex]
  & \texttt{PenDual} & 2.91 & 9.83 & 9.75 & 4.43 & 17.93 & 20.11 & 0.83 & 4.15 & 5.85 \\ [1.0ex]
  & \texttt{SpLCM} & \textbf{2.09} & 9.50 & 9.71 & \textbf{3.40} & 14.83 & 16.45 & \textbf{0.80} & 3.36 & 5.25 \\ [1.0ex]
  \hdashline
  & \texttt{SpLCM(O)} & 1.23 & 7.33 & 2.58 & 2.42 & 12.18 & 9.77 & 0.75 & 2.83 & 3.08 \\ [1.0ex]
  \hline
\end{tabular}
}
\caption{Comparison of the average $\ell_2$ norm distance of the off-diagonal elements $\|\text{vech}(\wh{\bSigma} - \bSigma^\ast)_{[o]}\|_2$, the Frobenius norm distance $\|\wh{\bSigma} - \bSigma^\ast\|_F$ and the operator norm distance $\|\wh{\bSigma} - \bSigma^\ast\|_2$. Except for the values for \texttt{SpLCM(O)}, the values with the lowest distance have been bolded.
} 
\label{tb:total_cov}
\end{table}

\begin{table}[!t]
\centering
\scalebox{0.8}{
\begin{tabular}{llrrrrrr}
  \hline \hline
  &  & \multicolumn{3}{c}{\uline{\quad \quad \quad \quad \; TPR \; \quad \quad \quad \quad}} & \multicolumn{3}{c}{\uline{\quad \quad \quad \quad \; FPR \; \quad \quad \quad \quad}} \\[1.0ex]
  & method & MA(1) & Random & Hub & MA(1) & Random & Hub \\[1.0ex] \hline
  $n=100$ & \texttt{Soft} & \textbf{1.000} & \textbf{0.991} & 0.899 & 0.085 & 0.044 & 0.052   \\ [1.0ex]
  $p=50$ & \texttt{SpCov} & 0.991 & 0.943 & 0.978 & 0.144 & 0.032 & 0.132 \\ [1.0ex]
  & \texttt{ProxCov} & 0.995 & 0.952 & 0.973 & \textbf{0.017} & \textbf{0.009} & 0.039 \\ [1.0ex]
  & \texttt{PenDual} & \textbf{1.000} & 0.977 & 0.142 & 0.024 & 0.010 & \textbf{0.000} \\ [1.0ex]
  & \texttt{SpLCM} & \textbf{1.000} & 0.952 & \textbf{1.000} & 0.026 & 0.015 & 0.009 \\ [1.0ex]
  \hdashline
  & \texttt{SpLCM(O)} & 1.000 & 0.992 & 1.000 & 0.019 & 0.011 & 0.002 \\ [1.0ex]
  \hline
  $n=100$ & \texttt{Soft} & 0.999 & \textbf{0.893} & 0.399 & 0.046 & 0.041 & 0.015   \\ [1.0ex]
  $p=100$ & \texttt{SpCov} & 0.375 & 0.388 & 0.069 & 0.023 & 0.013 & 0.002 \\ [1.0ex]
  & \texttt{ProxCov} & \textbf{1.000} & 0.842 & 0.261 & 0.027 & 0.097 & 0.056 \\ [1.0ex]
  & \texttt{PenDual} & \textbf{1.000} & 0.586 & 0.399 & \textbf{0.012} & \textbf{0.003} & \textbf{0.001} \\ [1.0ex]
  & \texttt{SpLCM} & \textbf{1.000} & 0.678 & \textbf{0.770} & 0.014 & 0.011 & 0.004 \\ [1.0ex]
  \hdashline
  & \texttt{SpLCM(O)} & 1.000 & 0.943 & 1.000 & 0.012 & 0.012 & 0.001 \\ [1.0ex]
  \hline
  $n=50$ & \texttt{Soft} & 0.949 & 0.340 & 0.041 & 0.034 & 0.010 & 0.002   \\ [1.0ex]
  $p=100$ & \texttt{SpCov} & 0.060 & 0.033 & 0.004 & \textbf{0.002} & \textbf{0.000} & \textbf{0.000} \\ [1.0ex]
  & \texttt{ProxCov} & 0.936 & \textbf{0.544} & \textbf{0.312} & 0.137 & 0.170 & 0.131 \\ [1.0ex]
  & \texttt{PenDual} & 0.928 & 0.027 & 0.000 & 0.007 & \textbf{0.000} & \textbf{0.000} \\ [1.0ex]
  & \texttt{SpLCM} & \textbf{0.979} & 0.144 & 0.013 & 0.015 & 0.002 & \textbf{0.000} \\ [1.0ex]
  \hdashline
  & \texttt{SpLCM(O)} & 0.994 & 0.543 & 1.000 & 0.011 & 0.008 & 0.000 \\ [1.0ex]
  \hline
\end{tabular}
}
\caption{Comparison of true positive rate (TPR) and false positive rate (FPR) over 20 simulated datasets. Except for the values for \texttt{SpLCM(O)}, the values with the highest TPR and the lowest FPR have been bolded.
} 
\label{tb:tprfpr}
\end{table}

\section{Data Analysis}
\label{sec:data}

\subsection{Speech signal classification for Parkinson's disease}

We consider the quadratic discriminant analysis (QDA) setting to discriminate healthy individuals from the Parkinson's disease (PD) patients where the inverse covariance matrix for each group is a required input.
QDA is a classification technique which allows each class to have its own distinct covariance matrix.
In QDA with $G$ classes, each observation $\by_i$ is assigned to the $g$th class if $g \in \{1,\ldots,G\}$ maximizes the discriminant function as below
\begin{equation} \label{eq:qda}
\log| \wh{\bSigma}_g^{-1} | - (\by_i - \wh{\bmu}_g)^\top\wh{\bSigma}_g^{-1}(\by_i - \wh{\bmu}_g) + 2 \log(\hat{\pi}_g),
\end{equation}
where $\wh{\bmu}_g, \wh{\bSigma}_g, \hat{\pi}_g$ are the sample mean, the estimated covariance matrix and the sample proportion for the $g$th class, respectively. 

We consider the voice recording data \citep{little2009suitability} obtained from the UCI machine learning data repository, which was also used in \citet{rothman2012positive} for measuring the performance of their covariance matrix estimator.
The dataset comprises 195 voice samples $(n=195)$ from 31 individuals, of which 147 samples are from Parkinson's disease patients and 48 samples are from healthy controls. 
For each voice sample, a variety of features such as jitter and shimmer were extracted, capturing different aspects of dysphonia.
As in \citet{rothman2012positive}, we treated those 195 samples as independent samples.

To assess the classification performance, we adopted the approach in \citet{rothman2012positive}. In this approach, the dataset is partitioned into the training set and the test set, where the classification rule \eqref{eq:qda} is set up with the training set and then applied to the test set. Specifically, we considered two settings. In Setting 1, we copied the setting in \citet{rothman2012positive} where there are 65 training samples $(n_{train}=65)$ with 49 cases and 130 testing samples $(n_{test}=130)$ with 98 cases. In Setting 2, we flipped the sample size between the training set and the test set $(n_{train}=130; n_{test}=65)$. The classification performance was measured by the average misclassification rate for the test sets over 100 random partitions.
As in Section \ref{sec:simul}, we compared our proposed \texttt{SpLCM} with other competing methods \texttt{Soft}, \texttt{SpCov} and \texttt{ProxCov} as summarized in Table \ref{tb:data_PKS}. For Setting 1, \texttt{Soft} showed the misclassification rate of 0.219 and similar performance of \texttt{Soft} was also reported in \citet{rothman2012positive}. \texttt{SpCov} and \texttt{ProxCov} with misclassification rate of 0.201 and 0.162, respectively, improved over \texttt{Soft} but our proposed method \texttt{SpLCM} exhibited the lowest misclassification rate of 0.149. The superior performance of \texttt{SpLCM} was similarly observed in Setting 2 as well. 

\begin{table}[!t]
\centering
\scalebox{0.86}{
\begin{tabular}{lrrrr}
  \hline \hline
  Setting  & \quad \texttt{Soft} & \quad \texttt{SpCov} & \texttt{ProxCov} & \quad \texttt{SpLCM} \\[1.0ex] \hline
   Setting 1 & 0.219 & 0.201 & 0.162 & \textbf{0.149} \\ [1.0ex]
   Setting 2 & 0.223 & 0.198 & 0.159 & \textbf{0.140} \\ [1.0ex]
  \hline
\end{tabular}
}
\caption{Average misclassification rate over 100 random partitions of the voice recording data by \citet{little2009suitability}. The lowest misclassification rates have been bolded.
} 
\label{tb:data_PKS}
\end{table}

\subsection{Human gene expression data analysis}
One of the major tasks in genetic studies is the cluster analysis \citep{jiang2004cluster} which aims at identifying clusters of similar genes among numerous genes. Specifically, we consider the hierarchical clustering, where a correlation matrix is used as a measure of similarity between genes. 
We use a human gene expression dataset of 60 unrelated individuals of Northern and Western European ancestry from Utah (CEU), which were first analyzed by \citet{stranger2007population}. \citet{bhadra2013joint} selected 100 most variable probes out of the 47,293 total available probes corresponding to different Illumina TargetID, each corresponding to a different transcript and applied their Bayesian approach to estimate the inverse covariance matrix of the 100 genes. This dataset is publicly available in the \texttt{R} package \texttt{BDgraph} and, for details of this data, see \citet{bhadra2013joint}.

\begin{figure}[hbt!]
	\centering
	
	\scalebox{0.6}{
	\mbox{
		\subfigure{
			\begin{overpic}[width=4.9in,angle=0]
				{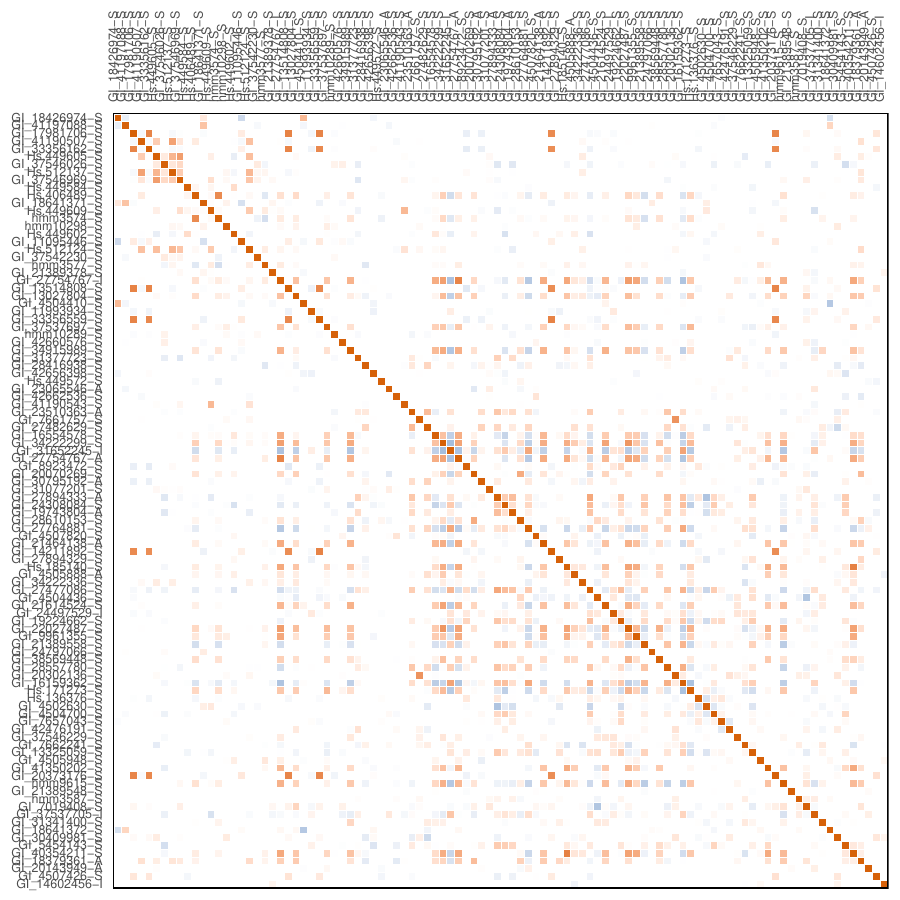}
            \put(50,105){\uline{\Large \texttt{Soft}}}
			\end{overpic}
		}
		\subfigure{
			\begin{overpic}[width=4.9in,angle=0]
				{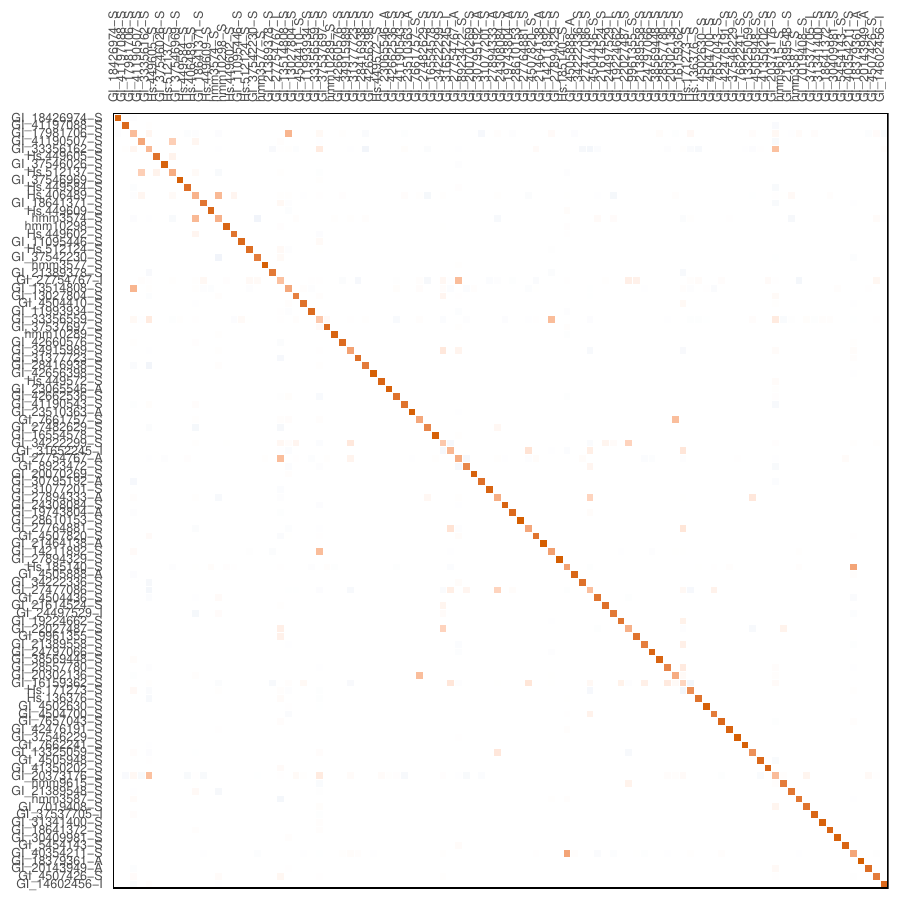}
            \put(48,105){\uline{\Large \texttt{SpCov}}}
			\end{overpic}
		}
	}
	
	}
	
	\vspace*{10mm}
	
	\scalebox{0.6}{
	\mbox{
		\subfigure{
			\begin{overpic}[width=4.9in,angle=0]
				{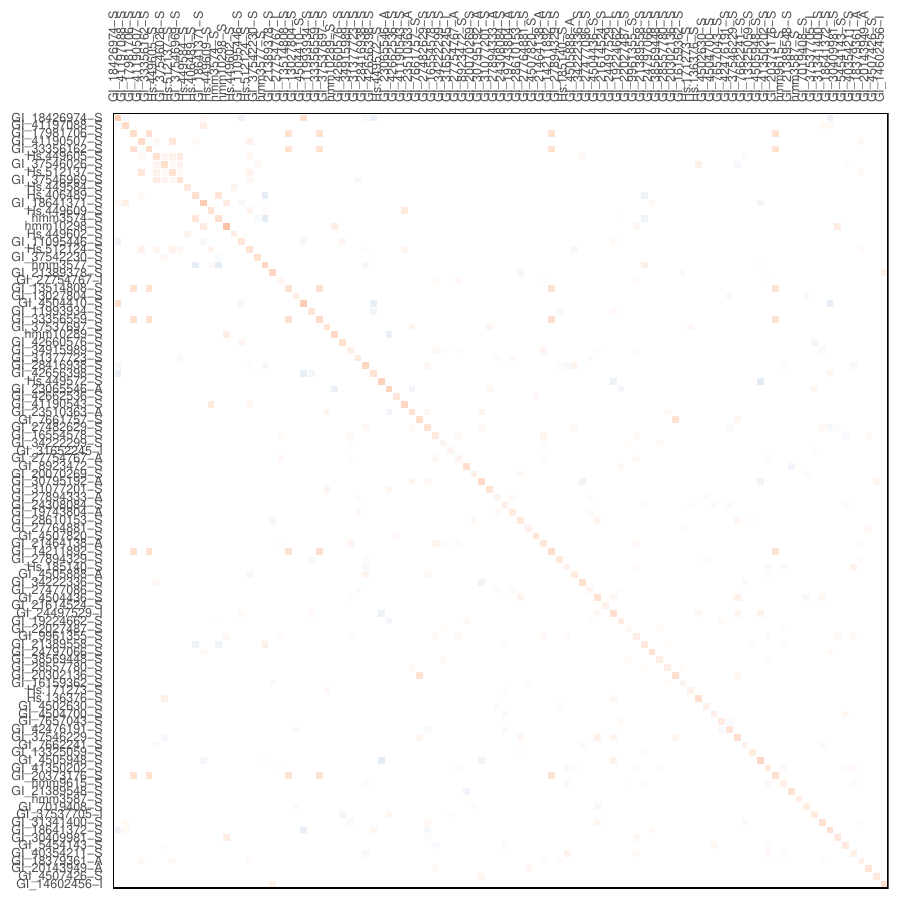}
            \put(46,105){\uline{\Large \texttt{ProxCov}}}
			\end{overpic}
		}
		\subfigure{
			\begin{overpic}[width=4.9in,angle=0]
				{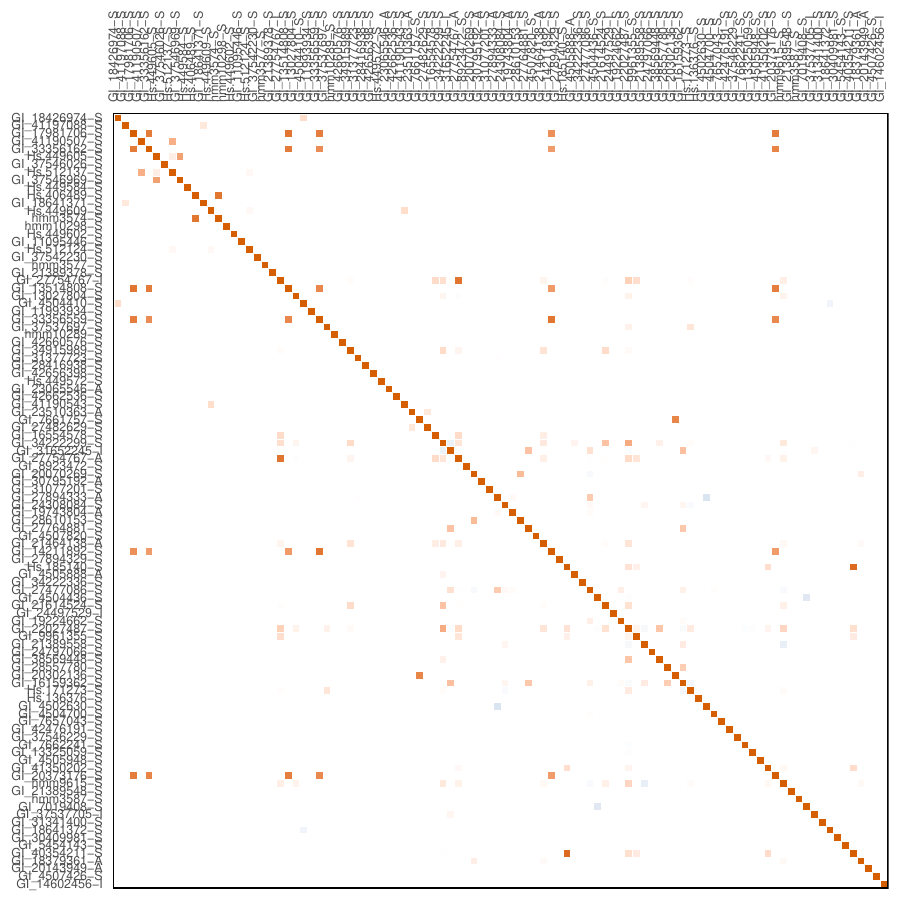}
            \put(48,105){\uline{\Large \texttt{SpLCM}}}
			\end{overpic}
		}
	}
	
	}
	

\caption{Heatmap of the covariance matrix by \texttt{Soft}, \texttt{SpCov}, \texttt{ProxCov} and \texttt{SpLCM}. Positive values are shown in red and negative values are shown in blue.}
\label{fig:heat}
\end{figure}

To obtain a similarity measure for the hierarchical clustering, we first estimate the correlation matrix by estimating the covariance matrix of the 100 scaled gene expressions for the 60 individuals $(n=60, p=100)$.
In Figure \ref{fig:heat}, the heatmaps of the covariance matrix estimators by \texttt{Soft}, \texttt{SpCov}, \texttt{ProxCov} and \texttt{SpLCM} are compared. Unlike \texttt{Soft} and \texttt{SpLCM}, the diagonal elements in \texttt{SpCov} and \texttt{ProxCov} are not fixed to one although we are using the scaled data, requiring additional adjustment to obtain the correlation matrix. Also, compared to other methods, \texttt{Soft} contains many non-zero elements, potentially leading to harder interpretation of the clustering result, as will be discussed. Also, such a non-sparse matrix estimator is less useful when the identification of correlation among genes is of interest \citep{butte2000discovering, zhang2005general, danaher2015covariance, wang2024wendy}.
On the other hand, \texttt{SpLCM} exhibits a clearer pattern of sparsity with diagonals fixed to one, giving a valid correlation matrix.

\begin{figure}[hbt!]
	\centering
	
	\scalebox{1.0}{
	\mbox{
		\subfigure{
			\begin{overpic}[width=4.9in,angle=0]
				{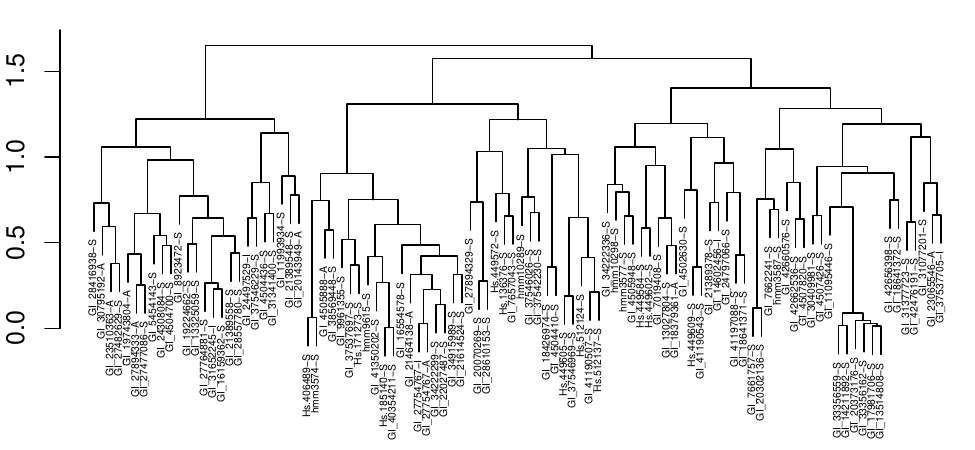}
			\end{overpic}
		}
		
	}
	
	}
	
	
	\scalebox{1.0}{
	\mbox{
		\subfigure{
			\begin{overpic}[width=4.9in,angle=0]
				{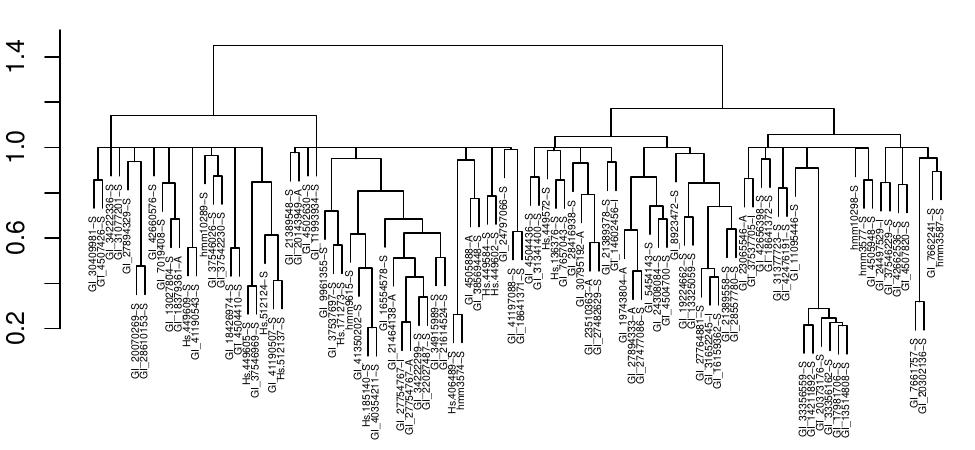}
			\end{overpic}
		}
		
	}
	
	}
	
	
	\scalebox{1.0}{
	\mbox{
		\subfigure{
			\begin{overpic}[width=4.9in,angle=0]
				{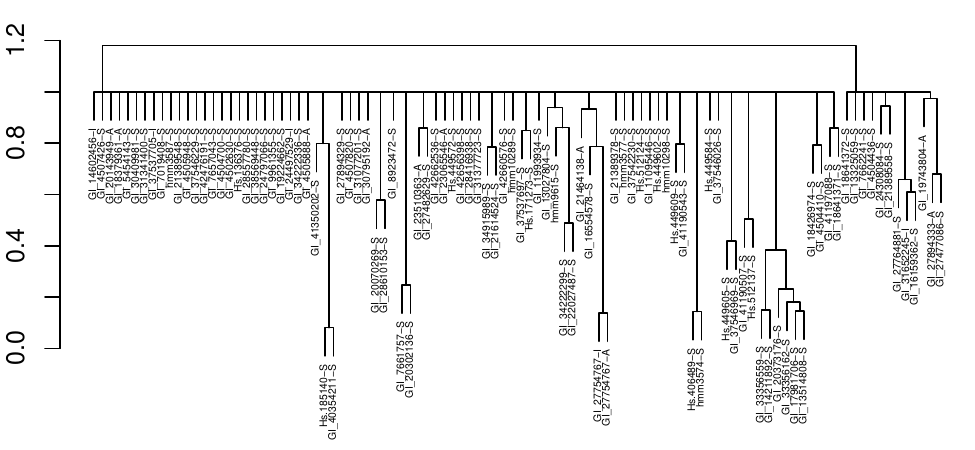}
            \put(12,38){A}
            \put(89,38){B}
			\end{overpic}
		}
		
	}
	
	}

\caption{Cluster dendrogram from the hierarchical clustering with the correlation matrix by \texttt{Sample} (top) \texttt{Soft} (middle) and \texttt{SpLCM} (bottom). In the bottom panel by \texttt{SpLCM}, the genes were clustered by two major branches, denoted as A and B with 15 clusters and 3 clusters, respectively.}
\label{fig:den}
\end{figure}

\begin{figure}[hbt!]
	\centering
	\scalebox{1.0}{
	\mbox{
		\subfigure{
			\begin{overpic}[width=6.0in,angle=0]
				{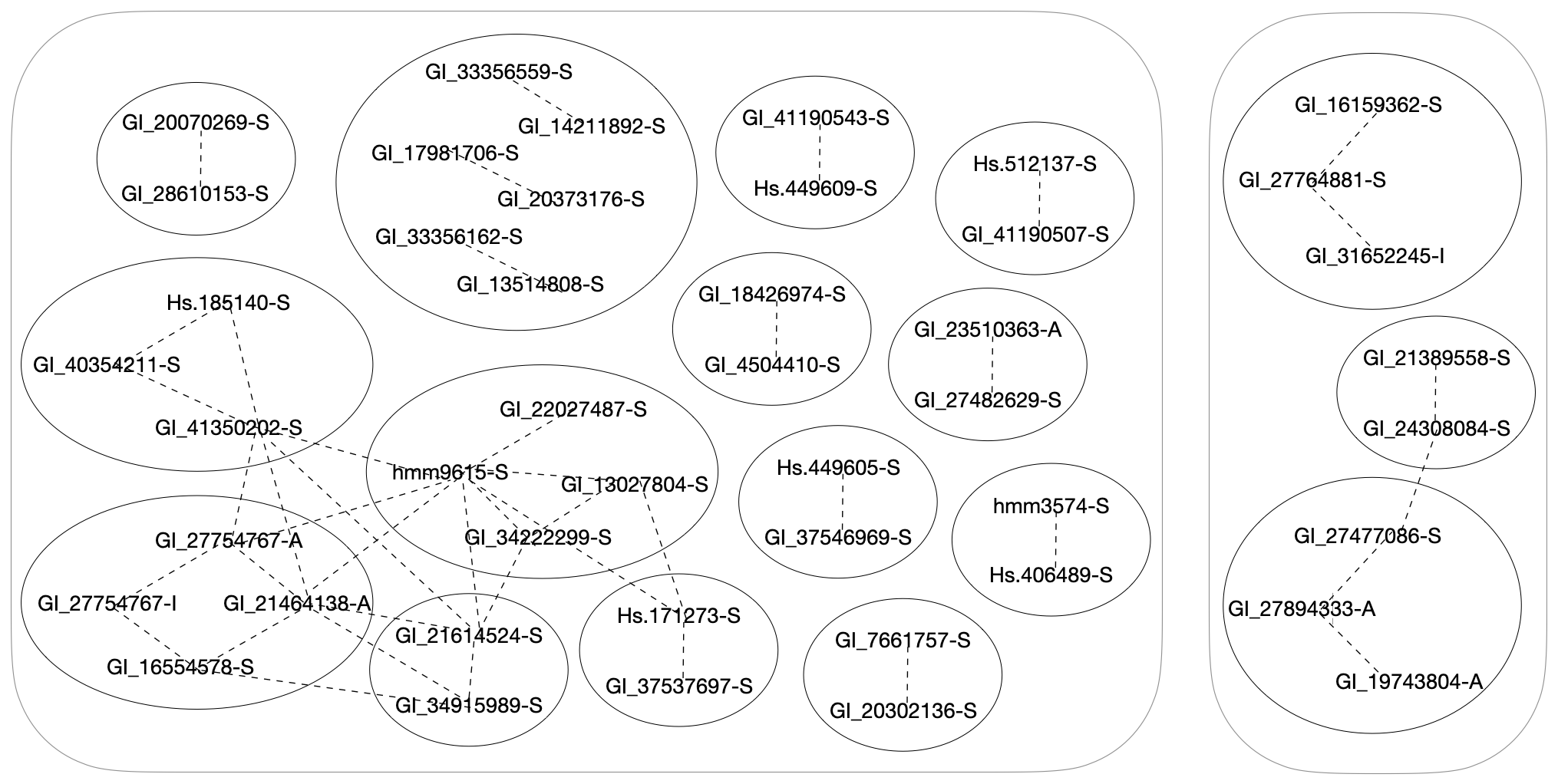}
            \put(38,50){A}
            \put(87,50){B}
			\end{overpic}
		}	
	}
	}
\caption{Clusters of genes identified by \texttt{SpLCM} shown in circles and the precision matrix graph by \citet{gu2015local} shown with dashed lines. The genes under the branches A and B in Figure \ref{fig:den} are contained in boxes A and B, respectively.}
\label{fig:clustergraph}
\end{figure}

Next, we applied our estimated correlation matrices to the hierarchical clustering problem where the 100 genes are clustered based on the similarity measured by the correlation.
In Figure \ref{fig:den}, we compare the cluster dendrograms from the hierarchical clustering where the applied correlation matrix is either \texttt{Sample}, \texttt{Soft} or \texttt{SpLCM}. Due to the sparsity in the correlation matrix, the dendrogram for \texttt{SpLCM} shows much simpler tree structure of the genes than that for \texttt{Sample} and \texttt{Soft}, demonstrating the improved interpretability.
In addition, many clusters identified in the dendrogram by \texttt{SpLCM} were also identified as conditionally independent to each other in the graphical model by \citet{gu2015local}. Specifically, the genes under the branch A and those under the branch B in the bottom panel of Figure \ref{fig:den} were not connected by any edges found by \citet{gu2015local} as shown in Figure \ref{fig:clustergraph}. Furthermore, the genes within each cluster by \texttt{SpLCM} were also connected by the graphical model by \citet{gu2015local} (i.e. edges within clusters) but many clusters were found to be conditionally independent with each other (i.e. edges between clusters). This is reasonable because the genes in different clusters might be conditionally independent. Indeed, if the covariance and the inverse covariance matrix take the block structure and each block consists of the variables in each cluster, similar results between the cluster analysis and the graphical model are expected.

\clearpage

\section{Conclusion}
\label{sec:conc}

In this paper, we formulate a linear covariance model that considers correlation among the elements in the sample covariance matrix and suggest a penalized regression for sparse estimation. 
We propose an ADMM algorithm to ensure positive definiteness of the estimated covariance matrix and establish theoretical properties of the proposed covariance matrix estimator. The superior performance of our proposed estimator is evidenced by simulation studies and data analysis.

Despite the superior performance of our proposed estimator, there is still a large gap between our proposed estimator with $\wh{\bV}^{-1}$ and the oracle sparse LCM estimator with $\bV^{-1}$, which necessitates future research on better estimating $\bV^{-1}$ before solving the optimization \eqref{eq:obj_covreg}.
In this paper, $\wh{\bV}^{-1}$ is constructed based on the CLIME estimator $\wh{\bOmega}$. Other estimators of $\bOmega^\ast$ such as the graphical LASSO may be considered for constructing $\wh{\bV}^{-1}$. In that case, theoretical properties of our proposed estimator 
need to be discussed under different assumptions on the parameter space depending on the choice of $\wh{\bOmega}$ and we leave this for future research.

When the normality of $\by$ cannot be assumed, our proposed approach to estimate $\bV^{-1}$ based on an estimator of $\bOmega^\ast$ is not applicable because the equation \eqref{eq:errorcov_gaussian} does not hold.
In this case, we may consider estimating $\bV$ or $\bV^{-1}$ directly, for example, via the bootstrap procedure as follows.
\begin{itemize}
    \item Generate $N$ bootstrap datasets where each dataset contains $n$ bootstrap samples.
    \item Compute the sample covariance matrix $\bS_b$ for each bootstrap dataset $(b=1,\ldots,N)$.
    \item Using $\text{vech}(\bS_1),\ldots,\text{vech}(\bS_N)$ as $N$ samples of $\text{vech}(\bS)$, compute the sample covariance matrix $\wh{\bV}^{boot}$ as an estimator of ${\rm Cov}\{\text{vech}(\bS)\}$.
\end{itemize}
However, such estimators of $\bV$ or $\bV^{-1}$ can be very noisy as the number of unknown parameters grows at the order of $p^4$. Alternatively, one may consider regularization approaches such as the CLIME estimation procedure over the whole $\bV^{-1}$ and we leave this for future research.

\bibliographystyle{agsm}

\bibliography{Main_arxiv}

\newpage
\renewcommand{\thesubsection}{S\arabic{subsection}}
\renewcommand{\theequation}{S\arabic{equation}}
\renewcommand{\thefigure}{S\arabic{figure}}
\renewcommand{\thetable}{S\arabic{table}}
\setcounter{equation}{0}
\setcounter{figure}{0}
\setcounter{table}{0}
\setcounter{page}{1}

\begin{center}
{\large\bf Supplementary Materials for ``Linear covariance model for sparse covariance matrices"} 
\medskip
\end{center}

\subsection{Relationship between the linear covariance model and thresholding estimators} \label{rel_lcm_thres}
We express $\bSigma$ using the linear covariance model by \citet{anderson1973asymptotically} as
\be \label{eq:linear_cov_supp}
\bSigma = \sum_{l=1}^{L}\sigma_l \bG_l
\ee
such that $\text{vech}(\bG_l)$ be the $l$th vector in the canonical basis of $\mathbb{R}^L$ where $L=p(p+1)/2$.
Let $\sigma_l$ and $s_l$ be the $l$th elements of $\text{vech}(\bSigma)$ and $\text{vech}(\bS)$, respectively,
and denote $\wh{\text{var}}(s_l)$ as the estimated variance of $s_l$. 
When $\wh{\bV}$ is a diagonal matrix with its $l$th diagonal element as $\wh{\text{var}}(s_l)$, we claim that the solution to \eqref{eq:obj_covreg} with $P(\bsigma_{[o]})=\|\bsigma_{[o]}\|_1$ leads to the adaptive soft thresholding estimator by \citet{cai2011adaptive}. The relation between the universal thresholding estimator and the solution to \eqref{eq:obj_covreg} with $\wh{\bV}=\bI_{p(p+1)/2}$ is trivial from this result. 

By differentiating the unconstrained objective function in \eqref{eq:obj_covreg} with $P(\bsigma_{[o]})=\|\bsigma_{[o]}\|_1$, the subgradient equation for $\bsigma$ is given by
\bse
- (\wh{\bV}^{-1} \bs - \wh{\bV}^{-1} \bsigma)/n + \lambda \bu = 0
\ese
where $\bu \in \mathbb{R}^L$ is a vector with the $l$th element as
\begin{align*}
u_l &= 0 \quad \quad \quad \quad \quad \text{if} \quad l \in \text{[d]}\\
&= \text{sign}(\sigma_l) \quad \quad \: \text{if} \quad l \in \text{[o]} \quad \text{and} \quad \sigma_l \neq 0\\
&\in [-1,1] \quad \quad \;\;\; \text{if} \quad l \in \text{[o]} \quad \text{and} \quad \sigma_l = 0.
\end{align*}
The above equation is equivalent to the following 
\begin{align*}
&s_l - \sigma_l = 0 \quad \quad \quad \quad \quad \quad \quad \quad \:\; \text{if} \quad l \in \text{[d]}\\
&s_l - \sigma_l - \lambda n \wh{\text{var}}(s_l) u_l = 0 \quad \quad \; \text{if} \quad l \in \text{[o]}
\end{align*}
which give the solution for each $\beta_l$ as
\begin{align*}
&\hat{\sigma}_l = s_l \quad \quad \quad \quad \quad \quad \quad \; \; \text{if} \quad l \in \text{[d]}\\
&\hat{\sigma}_l = S_{\lambda n \wh{\text{var}}(s_l)}(s_l) \quad \quad \quad \: \text{if} \quad l \in \text{[o]}
\end{align*}
where $S_{\lambda}(\cdot)$ is the soft thresholding operator at $\lambda$.
Note that the level of the threshold is determined by adapting the variability of individual elements of the sample covariance matrix $\bS$, as in the adaptive thresholding estimator proposed by \citet{cai2011adaptive}.

\subsection{Relationship between the linear covariance model and penalized likelihood estimators} \label{rel_lcm_penal}
Consider the model \eqref{eq:linear_cov_supp} and let $\ell(\bsigma)$ be the penalized likelihood for $\bSigma$ 
\bse
\ell(\bsigma) = \log \det (\bSigma) + \text{tr}(\bSigma^{-1} \bS) + \lambda \|\bP \circ \bSigma\|_1.
\ese
Differentiating $\ell(\bsigma)$ with respect to $\sigma_l$ gives
\begin{align*}
\frac{\partial \ell}{\partial \sigma_l} &= \text{tr}\bigg\{ \bigg(\frac{\partial \ell}{\partial \bSigma}\bigg)^\top \bigg(\frac{\partial \bSigma}{\partial \sigma_l}\bigg) \bigg\}\\
&= \text{tr}( \bSigma^{-1}\bG_l - \bSigma^{-1}\bS\bSigma^{-1}\bG_l  ) + 2 \lambda u_l
\end{align*}
where $u_l$ is as defined in Section \ref{rel_lcm_thres}. 
Hence, the estimating equation to minimize $\ell(\bsigma)$ is the system of equations
\bse
\text{tr}(\bSigma^{-1}\bG_l) = \text{tr}(\bSigma^{-1}\bS\bSigma^{-1}\bG_l) - 2 \lambda u_l
\ese
for all $l \in \{1,\ldots,L\}$. Since $\text{tr}(\bSigma^{-1}\bG_l)=\text{tr}(\bSigma^{-1}\bG_l\bSigma^{-1}\bSigma)=\text{tr}(\bSigma^{-1}\bG_l\bSigma^{-1}\sum_{m=1}^L\sigma_m \bG_m)$, the equation above can also be written as 
\be \label{eq:lcm_esteq}
\bC \bsigma = \bd - 2 \lambda \bu
\ee
where $\bu=(u_1,\ldots,u_L)^\top$ and
\bse
\bC = 
\begin{bmatrix}
\text{tr}(\bSigma^{-1}\bG_1 \bSigma^{-1}\bG_1) & \text{tr}(\bSigma^{-1}\bG_1 \bSigma^{-1}\bG_2) & ... & \text{tr}(\bSigma^{-1}\bG_1 \bSigma^{-1}\bG_L) \\
\text{tr}(\bSigma^{-1}\bG_2 \bSigma^{-1}\bG_1) & \text{tr}(\bSigma^{-1}\bG_2 \bSigma^{-1}\bG_2) & ... & \text{tr}(\bSigma^{-1}\bG_2 \bSigma^{-1}\bG_L) \\
\vdots & \vdots & \ddots & \vdots \\
\text{tr}(\bSigma^{-1}\bG_L \bSigma^{-1}\bG_1) & \text{tr}(\bSigma^{-1}\bG_L \bSigma^{-1}\bG_2) & ... & \text{tr}(\bSigma^{-1}\bG_L \bSigma^{-1}\bG_L)
\end{bmatrix}, \bd =
\begin{bmatrix}
\text{tr}(\bSigma^{-1}\bG_1 \bSigma^{-1}\bS) \\
\text{tr}(\bSigma^{-1}\bG_2 \bSigma^{-1}\bS) \\
\vdots \\
\text{tr}(\bSigma^{-1}\bG_L \bSigma^{-1}\bS)
\end{bmatrix}
\ese
as discussed in \citet{anderson1973asymptotically}.
Since 
\begin{align*}
\text{tr}(\bSigma^{-1}\bG_l \bSigma^{-1}\bG_m) &= \text{vec}(\bG_m)^\top (\bSigma^{-1} \otimes \bSigma^{-1}) \text{vec}(\bG_l)\\
&= \text{vech}(\bG_m)^\top \bD_p^\top (\bSigma^{-1} \otimes \bSigma^{-1}) \bD_p \text{vech}(\bG_l)
\end{align*}
and
\begin{align*}
\text{tr}(\bSigma^{-1}\bG_l \bSigma^{-1}\bS) &= \text{vec}(\bG_l)^\top (\bSigma^{-1} \otimes \bSigma^{-1}) \text{vec}(\bS)\\
&= \text{vech}(\bG_l)^\top \bD_p^\top (\bSigma^{-1} \otimes \bSigma^{-1}) \bD_p \text{vech}(\bS),
\end{align*}
using the fact that $\text{vech}(\bG_l)$ is the $l$th vector in the canonical basis of $\mathbb{R}^L$, the equation \eqref{eq:lcm_esteq} can be written as 
\bse
\bD_p^\top (\bSigma^{-1} \otimes \bSigma^{-1}) \bD_p \bsigma = \bD_p^\top (\bSigma^{-1} \otimes \bSigma^{-1}) \bD_p \bs - 2 \lambda \bu.
\ese
Since $\bD_p^\top (\bSigma^{-1} \otimes \bSigma^{-1}) \bD_p = \{\bD_p^+ (\bSigma \otimes \bSigma) \bD_p^{+\top}\}^{-1} = (2/n) \bV^{-1}$, we have
\bse 
\bV^{-1} \bsigma = \bV^{-1} \bs - \lambda n \bu,
\ese
which is the estimating equation for the penalized least squares problem
\bse
\argmin_{\bsigma} \frac{1}{2n} (\bs - \bsigma)^\top \bV^{-1} (\bs - \bsigma) + \lambda \sum_{l \in [o]} |\sigma_l| 
\ese
with fixed $\bV$. 
Since the problem above is equivalent to the optimization \eqref{eq:obj_covreg} without the diagonal constraint, the penalized likelihood estimator from \eqref{eq:log-lik} is the solution to the unconstrained sparse linear covariance model with known $\bV$.

\subsection{Proof of Proposition \ref{prop1}} \label{prop1_proof}
For any $\bsigma \in \mathbb{R}^{p(p+1)/2}$ such that $\bsigma_{[d]}=\bs_{[d]}$, the basic inequality gives
\bse
\frac{1}{2n}\|\bV^{-\frac{1}{2}} \bs-\bV^{-\frac{1}{2}} \wt{\bsigma}\|_2^2 + \lambda \|\wt{\bsigma}_{[o]}\|_1 \leq \frac{1}{2n}\|\bV^{-\frac{1}{2}} \bs-\bV^{-\frac{1}{2}} \bsigma\|_2^2 + \lambda \|\bsigma_{[o]}\|_1.
\ese
Since $\wt{\bsigma}_{[d]}=\bsigma_{[d]}=\bs_{[d]}$, the above inequality can be rewritten as
\bse 
\frac{1}{2n}\|\bV_o^{-\frac{1}{2}} \bs_{[o]} -\bV_o^{-\frac{1}{2}} \wt{\bsigma}_{[o]}\|_2^2 + \lambda \|\wt{\bsigma}_{[o]}\|_1 \leq \frac{1}{2n}\|\bV_o^{-\frac{1}{2}} \bs_{[o]}-\bV^{-\frac{1}{2}} \bsigma_{[o]}\|_2^2 + \lambda \|\bsigma_{[o]}\|_1.
\ese

Since $\bs_{[o]}={\bsigma^\ast}_{[o]}+\bepsilon_{[o]}$ by \eqref{eq:linear_covreg}, we have
\bse
\frac{1}{2n}\|\bV_o^{-\frac{1}{2}} ({\bsigma^\ast}_{[o]} - \wt{\bsigma}_{[o]}) + \bV_o^{-\frac{1}{2}} \bepsilon_{[o]}\|_2^2 + \lambda \|\wt{\bsigma}_{[o]}\|_1 \leq \frac{1}{2n}\|\bV_o^{-\frac{1}{2}} ({\bsigma^\ast}_{[o]} - \bsigma_{[o]}) + \bV_o^{-\frac{1}{2}} \bepsilon_{[o]}\|_2^2 + \lambda \|\bsigma_{[o]}\|_1.
\ese
Rearranging the terms above gives
\begin{align*}
\frac{\|\bV_o^{-\frac{1}{2}} ({\bsigma^\ast}_{[o]} - \wt{\bsigma}_{[o]})\|_2^2}{2n} &\leq \frac{\|\bV_o^{-\frac{1}{2}} ({\bsigma^\ast}_{[o]} - \bsigma_{[o]})\|_2^2}{2n} + \frac{ (\wt{\bsigma}_{[o]}-\bsigma_{[o]})^\top (\bV_o^{-\frac{1}{2}})^\top \bV_o^{-\frac{1}{2}} \bepsilon_{[o]} }{n} + \lambda(\|\bsigma_{[o]}\|_1-\|\wt{\bsigma}_{[o]}\|_1)\\
&\leq \frac{\|\bV_o^{-\frac{1}{2}} ({\bsigma^\ast}_{[o]} - \bsigma_{[o]})\|_2^2}{2n} + \frac{ \|\wt{\bsigma}_{[o]}-\bsigma_{[o]}\|_1 \|(\bV_o^{-\frac{1}{2}})^\top \bV_o^{-\frac{1}{2}} \bepsilon_{[o]}\|_\infty}{n} + \lambda(\|\bsigma_{[o]}\|_1-\|\wt{\bsigma}_{[o]}\|_1)\\
&\leq \frac{\|\bV_o^{-\frac{1}{2}} ({\bsigma^\ast}_{[o]} - \bsigma_{[o]})\|_2^2}{2n} + \frac{ \lambda \|\wt{\bsigma}_{[o]}-\bsigma_{[o]}\|_1}{2} + \lambda(\|\bsigma_{[o]}\|_1-\|\wt{\bsigma}_{[o]}\|_1),
\end{align*}
where the second inquality holds by H\"{o}lder's inequality and the third inequality holds because $\lambda \geq 2\|(\bV_o^{-\frac{1}{2}})^\top \bV_o^{-\frac{1}{2}} \bepsilon_{[o]}\|_\infty/n$ is assumed. 

For $\bsigma_{[o]} \in \mathbb{R}^{p(p-1)/2}$, denote $(\bsigma_{[o]})_\mathcal{S} \in \mathbb{R}^{|\mathcal{S}|}$ as the sub-vector of $\bsigma_{[o]}$ indexed by $\mathcal{S}$ and denote $(\bsigma_{[o]})_{\mathcal{S}^c}$ similarly for $\mathcal{S}^c$. 
By setting $\bsigma_{[o]}={\bsigma^\ast}_{[o]}$,
\begin{align*} 
\frac{\|\bV_o^{-\frac{1}{2}} ({\bsigma^\ast}_{[o]} - \wt{\bsigma}_{[o]})\|_2^2}{2n} & \leq \frac{\lambda}{2} \|{\bsigma^\ast}_{[o]}-\wt{\bsigma}_{[o]}\|_1 + \lambda \|\bsigma_{[o]}^\ast\|_1 - \lambda \|\wt{\bsigma}_{[o]}\|_1 \nonumber \\
& \leq \frac{\lambda}{2} \|{\bsigma^\ast}_{[o]}-\wt{\bsigma}_{[o]}\|_1 + \lambda \|(\bsigma_{[o]}^\ast-\wt{\bsigma}_{[o]})_\mathcal{S}\|_1 - \lambda \|(\bsigma_{[o]}^\ast-\wt{\bsigma}_{[o]})_{\mathcal{S}^c}\|_1 \nonumber \\
& \leq \frac{3\lambda}{2} \|(\bsigma_{[o]}^\ast-\wt{\bsigma}_{[o]})_\mathcal{S}\|_1 - \frac{\lambda}{2} \|(\bsigma_{[o]}^\ast-\wt{\bsigma}_{[o]})_{\mathcal{S}^c}\|_1\\
& \leq \frac{3\lambda \sqrt{s}}{2} \|\bsigma_{[o]}^\ast-\wt{\bsigma}_{[o]}\|_2 \nonumber
\end{align*}
where the second inequality holds because
\begin{align*}
\lambda \|\bsigma_{[o]}^\ast\|_1 - \lambda \|\wt{\bsigma}_{[o]}\|_1 & = \lambda \|(\bsigma_{[o]}^\ast)_\mathcal{S}\|_1 - \lambda \|\bsigma_{[o]}^\ast + \wt{\bsigma}_{[o]} - \bsigma_{[o]}^\ast\|_1\\
& = \lambda \|(\bsigma_{[o]}^\ast)_\mathcal{S}\|_1 - \lambda \|(\bsigma_{[o]}^\ast + \wt{\bsigma}_{[o]} - \bsigma_{[o]}^\ast)_\mathcal{S}\|_1 - \lambda \|(\wt{\bsigma}_{[o]} - \bsigma_{[o]}^\ast)_{\mathcal{S}^c}\|_1\\
& \leq \lambda \|(\bsigma_{[o]}^\ast)_\mathcal{S}\|_1 - \lambda \{ \|(\bsigma_{[o]}^\ast)_\mathcal{S}\|_1 - \| (\wt{\bsigma}_{[o]} - \bsigma_{[o]}^\ast)_\mathcal{S}\|_1\} - \lambda \|(\wt{\bsigma}_{[o]} - \bsigma_{[o]}^\ast)_{\mathcal{S}^c}\|_1\\
& = \lambda \|(\bsigma_{[o]}^\ast-\wt{\bsigma}_{[o]})_\mathcal{S}\|_1 - \lambda \|(\bsigma_{[o]}^\ast-\wt{\bsigma}_{[o]})_{\mathcal{S}^c}\|_1.
\end{align*}

By Assumption \ref{ass:eigen_bound}, we have
\bse
\frac{({\bsigma^{\ast}}_{[o]}-\wt{\bsigma}_{[o]})^\top (\bV_o^{-\frac{1}{2}})^\top (\bV_o^{-\frac{1}{2}}) ({\bsigma^{\ast}}_{[o]}-\wt{\bsigma}_{[o]})}{n \|{\bsigma^{\ast}}_{[o]}-\wt{\bsigma}_{[o]}\|_2^2} \geq \frac{1}{C^2}
\ese
because $(\bV_o^{-\frac{1}{2}})^\top (\bV_o^{-\frac{1}{2}})/n$ is a sub-matrix of $n^{-1}\bV^{-1}$, which has the smallest eigenvalue greater than $C^{-2}$.
Hence, we have
\bse
\frac{1}{2 C^2} \|{\bsigma^{\ast}}_{[o]}-\wt{\bsigma}_{[o]}\|_2^2 \leq \frac{3\lambda \sqrt{s}}{2} \|{\bsigma^{\ast}}_{[o]}-\wt{\bsigma}_{[o]}\|_2,
\ese
which gives
\bse
\|{\bsigma^{\ast}}_{[o]}-\wt{\bsigma}_{[o]}\|_2 \leq 3\lambda C^2 \sqrt{s}.
\ese
\qed

\subsection{Proof of Theorem \ref{thm1}} \label{thm1_proof}

We first state the following lemma.
\begin{Lem} \label{lem1}
For all $i=1,\ldots,n$ and $l=1,\ldots,p(p-1)/2$, $\xi_{il}$ is sub-exponential with a uniform bound $K_1=\max_{i,l} \| \xi_{il} \|_{\psi_1}$. 
\end{Lem}

To prove Theorem \ref{thm1}, it is sufficient if we show that  $\lambda \geq 2\|(\bV_o^{-\frac{1}{2}})^\top \bV_o^{-\frac{1}{2}} \bepsilon_{[o]}\|_\infty/n$ with probability at least $1- p^{2-c_1 C_1^2}$. 
Define the random event $\mathcal{A}$ as
\bse
\mathcal{A} = \{ 2\|(\bV_o^{-\frac{1}{2}})^\top \bV_o^{-\frac{1}{2}} \bepsilon_{[o]}\|_\infty/n < \lambda \}.
\ese

Since $\bepsilon_{[o]} = \bs_{[o]} - \bsigma^\ast_{[o]} = \sum_{i=1}^n n^{-1} \text{vech}(\by_i \by_i^\top - \bSigma^\ast)_{[o]}$, 
\begin{align*}
Pr(\mathcal{A}^c) 
& = Pr\bigg(\|(\bV_o^{-\frac{1}{2}})^\top \bV_o^{-\frac{1}{2}} \bepsilon_{[o]}\|_\infty \geq \frac{\lambda n}{2}\bigg)\\
& = Pr\bigg(\| \sum_{i=1}^n n^{-1} (\bV_o^{-\frac{1}{2}})^\top \bV_o^{-\frac{1}{2}} \text{vech}(\by_i \by_i^\top - \bSigma^\ast)_{[o]}\|_\infty \geq \frac{\lambda n}{2}\bigg)\\
& = Pr\bigg\{ \bigcup_{l=1}^{p(p-1)/2} \bigg( | \sum_{i=1}^n \xi_{il} | \geq \frac{\lambda n}{2}  \bigg) \bigg\}\\
& \leq \sum_{l=1}^{p(p-1)/2} Pr \bigg( | \sum_{i=1}^n \xi_{il} | \geq \frac{\lambda n}{2}  \bigg)\\
& \leq p(p-1) \exp \bigg\{ - c_1 \min \bigg(\frac{\lambda^2 n^2 }{4 K_1^2 n}, \frac{\lambda n}{2K_1} \bigg) \bigg\}
\end{align*}
where the last inequality holds by Bernstein’s inequality (Theorem 2.8.2 of \citet{vershynin2018high}) and Lemma \ref{lem1}.
Since $\lambda \rightarrow 0$ as $n$ and $p$ increases, 
\begin{align*}
Pr(\mathcal{A}^c) 
& \leq p(p-1) \exp \bigg( -c_1\frac{\lambda^2 n}{4 K_1^2} \bigg)\\
& \leq p^2 \exp ( -c_1 C_1^2 \log p )\\
& \leq p^{2-c_1 C_1^2}.
\end{align*}
Hence, 
\bse
Pr(\mathcal{A}) = 1-Pr(\mathcal{A}^c) \geq 1- p^{2-c_1 C_1^2}.
\ese
\qed

\subsection{Proof of Theorem \ref{thm2}} \label{thm2_proof}
First, we state the following key lemmas.

\begin{Lem} \label{lem:Vmax_bound}
$n^{-1} \|\wh{\bV}^{-1}-\bV^{-1}\|_{\max} \leq 8 C_0 M^2/c \sqrt{\log p /n}$ with probability $1-4p^{-\tau}$ where $\tau>0$ and $C_0$ is as defined in \citet{cai2011constrained}.
\end{Lem}

\begin{Lem} \label{lem:re_cond_est}
For any non-zero vector $\balpha \in \mathbb{R}^{p(p-1)/2}$, denote $\balpha_{\mathcal{S}}$ as the sub-vector of $\balpha$ indexed by $\mathcal{S}$ and denote $\balpha_{\mathcal{S}^c}$ similarly for $\mathcal{S}^c$ where $\mathcal{S}$ and $s$ are as defined in Proposition \ref{prop1}.
For $\tau>0$ and $\balpha$ such that $\| \balpha_{\mathcal{S}^c} \|_1 \leq 3 \| \balpha_{\mathcal{S}} \|_1$, if 
\bse
s \leq \frac{c}{256 C_0 C^2 M^2} \sqrt{\frac{n}{\log p}}
\ese
where $C_0$ is as defined in \citet{cai2011constrained}, we have
\bse
\frac{ \| (\wh{\bV}_o^{-\frac{1}{2}}) \balpha \|_2^2 }{n \| \balpha \|_2^2} \geq \frac{1}{2C^2} 
\ese
with probability at least $1-4p^{-\tau}$.
\end{Lem}

Similar to the proof of Proposition \ref{prop1}, if $\lambda \geq 2 \|(\wh{\bV}_o^{-\frac{1}{2}})^\top \wh{\bV}_o^{-\frac{1}{2}} \bepsilon_{[o]}\|_\infty/n$, we can show that
\begin{equation} \label{eq:supeq1}
\frac{\|\wh{\bV}_o^{-\frac{1}{2}} ({\bsigma^\ast}_{[o]} - \wh{\bsigma}_{[o]})\|_2^2}{2n} \leq \frac{3\lambda}{2} \|(\bsigma_{[o]}^\ast-\wh{\bsigma}_{[o]})_\mathcal{S}\|_1 - \frac{\lambda}{2} \|(\bsigma_{[o]}^\ast-\wh{\bsigma}_{[o]})_{\mathcal{S}^c}\|_1,
\end{equation}
which leads to
\bse
\|(\bsigma_{[o]}^\ast-\wh{\bsigma}_{[o]})_{\mathcal{S}^c}\|_1 \leq 3 \|(\bsigma_{[o]}^\ast-\wh{\bsigma}_{[o]})_\mathcal{S}\|_1.
\ese
That is, $\bsigma_{[o]}^\ast-\wh{\bsigma}_{[o]}$ satisfies the condition for $\balpha$ in Lemma \ref{lem:re_cond_est} so that we have
\bse
\frac{({\bsigma^{\ast}}_{[o]}-\wh{\bsigma}_{[o]})^\top (\wh{\bV}_o^{-\frac{1}{2}})^\top (\wh{\bV}_o^{-\frac{1}{2}}) ({\bsigma^{\ast}}_{[o]}-\wh{\bsigma}_{[o]})}{n \|{\bsigma^{\ast}}_{[o]}-\wh{\bsigma}_{[o]}\|_2^2} \geq \frac{1}{2C^2} 
\ese
with probability $1-4p^{-\tau}$.
Hence, if $\lambda \geq 2 \|(\wh{\bV}_o^{-\frac{1}{2}})^\top \wh{\bV}_o^{-\frac{1}{2}} \bepsilon_{[o]}\|_\infty/n$, then, with probability $1-4p^{-\tau}$,
\bse
\frac{1}{4C^2} \|{\bsigma^{\ast}}_{[o]}-\wh{\bsigma}_{[o]}\|_2^2 \leq \frac{3\lambda \sqrt{s}}{2} \|{\bsigma^{\ast}}_{[o]}-\wh{\bsigma}_{[o]}\|_2,
\ese
which gives
\be \label{eq:error_bound_thm2}
\|{\bsigma^{\ast}}_{[o]}-\wh{\bsigma}_{[o]}\|_2 \leq 6\lambda C^2 \sqrt{s}. 
\ee
Hence, it is sufficient if we show that there exists a constant $C^\ast$ such that $\lambda = C^\ast \sqrt{\log p/n} \geq 2 n^{-1}\|(\wh{\bV}_o^{-\frac{1}{2}})^\top \wh{\bV}_o^{-\frac{1}{2}} \bepsilon_{[o]}\|_\infty$. Since
\bse
n^{-1}\|(\wh{\bV}_o^{-\frac{1}{2}})^\top \wh{\bV}_o^{-\frac{1}{2}} \bepsilon_{[o]}\|_\infty \leq n^{-1}\|(\bV_o^{-\frac{1}{2}})^\top \bV_o^{-\frac{1}{2}} \bepsilon_{[o]}\|_\infty + n^{-1}\|\{(\wh{\bV}_o^{-\frac{1}{2}})^\top \wh{\bV}_o^{-\frac{1}{2}} - (\bV_o^{-\frac{1}{2}})^\top \bV_o^{-\frac{1}{2}} \} \bepsilon_{[o]}\|_\infty
\ese
and the first term on the right hand has been bounded by $\sqrt{\log p/n}$ up to a constant with probability $1- p^{2-c_1 C_1^2}$ in the proof of Theorem \ref{thm1},
we will find a bound for $n^{-1}\|\{(\wh{\bV}_o^{-\frac{1}{2}})^\top \wh{\bV}_o^{-\frac{1}{2}} - (\bV_o^{-\frac{1}{2}})^\top \bV_o^{-\frac{1}{2}} \} \bepsilon_{[o]}\|_\infty$.

Consider the inequality
\bse
n^{-1}\|\{(\wh{\bV}_o^{-\frac{1}{2}})^\top \wh{\bV}_o^{-\frac{1}{2}} - (\bV_o^{-\frac{1}{2}})^\top \bV_o^{-\frac{1}{2}} \} \bepsilon_{[o]}\|_\infty \leq n^{-1}\| (\wh{\bV}_o^{-\frac{1}{2}})^\top \wh{\bV}_o^{-\frac{1}{2}} - (\bV_o^{-\frac{1}{2}})^\top \bV_o^{-\frac{1}{2}} \|_{1,\infty} \| \bepsilon_{[o]} \|_\infty.
\ese
Since the number of non-zero in each row of $(\bSigma^\ast)^{-1} \otimes (\bSigma^\ast)^{-1}$ is less than $t^2$ by Assumption \ref{ass:invcov_sparsity}, so is $\wh{\bOmega} \otimes \wh{\bOmega} - (\bSigma^\ast)^{-1} \otimes (\bSigma^\ast)^{-1}$ with probability $1-4p^{-\tau}$
by the model selection consistency of $\wh{\bOmega}$ \citep{cai2011constrained}. Hence,
\begin{align*}
n^{-1}\| (\wh{\bV}_o^{-\frac{1}{2}})^\top \wh{\bV}_o^{-\frac{1}{2}} - (\bV_o^{-\frac{1}{2}})^\top \bV_o^{-\frac{1}{2}} \|_{1,\infty} &\leq \| \wh{\bOmega} \otimes \wh{\bOmega} - (\bSigma^\ast)^{-1} \otimes (\bSigma^\ast)^{-1}  \|_{1,\infty}\\
&\leq t^2 o\bigg(\sqrt{\frac{\log p}{n}}\bigg)
\end{align*}
by Lemma \ref{lem:Vmax_bound}. Since we can show that $\| \bepsilon_{[o]} \|_\infty = o(\sqrt{\log p /n})$, we have shown that
\bse
n^{-1}\|\{(\wh{\bV}_o^{-\frac{1}{2}})^\top \wh{\bV}_o^{-\frac{1}{2}} - (\bV_o^{-\frac{1}{2}})^\top \bV_o^{-\frac{1}{2}} \} \bepsilon_{[o]}\|_\infty = t^2 o\bigg(\frac{\log p}{n}\bigg)
\ese
which converges as long as the order of $t^2$ is smaller than $n/\log p$. Particularly, if $t^2=o(\sqrt{n/\log p})$, 
$n^{-1}\|\{(\wh{\bV}_o^{-\frac{1}{2}})^\top \wh{\bV}_o^{-\frac{1}{2}} - (\bV_o^{-\frac{1}{2}})^\top \bV_o^{-\frac{1}{2}} \} \bepsilon_{[o]}\|_\infty=o(\sqrt{\log p/n})$.
Combining this result with the result of Theorem \ref{thm1}, we have shown that there exists a constant $c^\ast$ such that 
\begin{align*}
n^{-1}\|(\wh{\bV}_o^{-\frac{1}{2}})^\top \wh{\bV}_o^{-\frac{1}{2}} \bepsilon_{[o]}\|_\infty &\leq n^{-1}\|(\bV_o^{-\frac{1}{2}})^\top \bV_o^{-\frac{1}{2}} \bepsilon_{[o]}\|_\infty + n^{-1}\|\{(\wh{\bV}_o^{-\frac{1}{2}})^\top \wh{\bV}_o^{-\frac{1}{2}} - (\bV_o^{-\frac{1}{2}})^\top \bV_o^{-\frac{1}{2}} \} \bepsilon_{[o]}\|_\infty\\
&\leq c^\ast \sqrt{\frac{\log p}{n}}
\end{align*}
with probability $(1- p^{2-c_1 C_1^2}) (1-4p^{-\tau}) \geq 1 - 8 p^{\min(2-c_1 C_1^2, -\tau)}$. Hence, plugging $\lambda=c^\ast \sqrt{\log p/n}$ in \eqref{eq:error_bound_thm2}, we have
\bse
\|{\bsigma^{\ast}}_{[o]}-\wh{\bsigma}_{[o]}\|_2 \leq C^\ast \sqrt{\frac{s\log p}{n}}
\ese
where $C^\ast=6 c^\ast C^2$.
\qed

\subsection{Proof of Theorem \ref{thm3}} \label{thm13_proof}

We first state the following lemma.
\begin{Lem} \label{lem4}
Suppose assumptions in Theorem \ref{thm2} hold. Then, we have
\bse
\|\wh{\bsigma}_{[o]} - {\bsigma^\ast}_{[o]}\|_1 = O_P \bigg( s \sqrt{\frac{\log p}{n}} \bigg)
\ese
\end{Lem}

To prove Theorem \ref{thm3}, we use the fact that the operator norm of a symmetric matrix is bounded by the $\ell_1$ or $\ell_\infty$ norm of the matrix \citep{golub2013matrix}, that is,
\begin{align*}
\|\bSigma(\wh{\bsigma})-\bSigma^\ast\|_2 &\leq \max_{j} \sum_{k} |\bSigma(\wh{\bsigma})_{jk} - \bSigma_{jk}^\ast|
\end{align*}
where $\bSigma(\wh{\bsigma})_{jk}$ and $\bSigma_{jk}^\ast$ denote the $(j,k)$th element of $\bSigma(\wh{\bsigma})$ and $\bSigma^\ast$, respectively.
Since
\begin{align*}
\max_{j} \sum_{k} |\bSigma(\wh{\bsigma})_{jk} - \bSigma_{jk}^\ast| &= \max_{j} \bigg\{ |\bSigma(\wh{\bsigma})_{jj} - \bSigma_{jj}^\ast| + \sum_{k \neq j} |\bSigma(\wh{\bsigma})_{jk} - \bSigma_{jk}^\ast| \bigg\}\\
&\leq \max_{j}|s_{jj} - \bSigma_{jj}^\ast| + \|\wh{\bsigma}_{[o]} - {\bsigma^\ast}_{[o]}\|_1\\
&= O_p \bigg( \sqrt{\frac{\log p}{n}} \bigg) + O_p \bigg( s \sqrt{\frac{\log p}{n}} \bigg)
\end{align*}
where the last equality holds by Lemma \ref{lem4}. Hence, 
\bse
\|\bSigma(\wh{\bsigma})-\bSigma^\ast\|_2 = O_p \bigg( s \sqrt{\frac{\log p}{n}} \bigg).
\ese
\qed


\subsection{Proof of Lemma \ref{lem1}} \label{lem1_proof}
First, we show that each element of $n^{-1}\bV^{-1} \text{vech}(\by_i \by_i^\top - \bSigma^\ast)\in \mathbb{R}^{p(p+1)/2}$ is sub-exponential. Since
\begin{align*}
n^{-1}\bV^{-1} \text{vech}(\by_i \by_i^\top - \bSigma^\ast)&=2^{-1}\bD_p^\top ({\bSigma^\ast}^{-1}\otimes {\bSigma^\ast}^{-1}) \bD_p \text{vech}(\by_i \by_i^\top - \bSigma^\ast)\\
&=2^{-1}\bD_p^\top \text{vec}\{{\bSigma^\ast}^{-1} (\by_i \by_i^\top - \bSigma^\ast) {\bSigma^\ast}^{-1}\},
\end{align*}
it is sufficient if we prove that each element of ${\bSigma^\ast}^{-1} (\by_i \by_i^\top - \bSigma^\ast) {\bSigma^\ast}^{-1}$ is sub-exponential.
Denote $\bx_i={\bSigma^\ast}^{-1} \by_i$ and let $x_{ij}$ and $x_{ik}$ be the $j$th and the $k$th elements of $\bx_i$, respectively. 
Note that $x_{ij}x_{ik}$ is the $(j,k)$th element of ${\bSigma^\ast}^{-1} \by_i \by_i^\top {\bSigma^\ast}^{-1}$.
Since $\by_i \sim \mathbb{N}_p(\bzero, \bSigma^\ast)$, we have $\bx_i \sim \mathbb{N}_p(\bzero, {\bSigma^\ast}^{-1})$. Hence, $x_{ij}x_{ik}$ is the product of two normal random variables and, by Lemma 2.7.7 of \citet{vershynin2018high}, we can show that the $(j,k)$th element $({\bSigma^\ast}^{-1} \by_i \by_i^\top {\bSigma^\ast}^{-1})_{jk}$ is sub-exponential with its sub-exponential norm bounded by $c^{-2}$ up to a constant  across all $j$ and $k$. Also, since $\mathbb{E}({\bSigma^\ast}^{-1} \by_i \by_i^\top {\bSigma^\ast}^{-1})={\bSigma^\ast}^{-1}$, each element of ${\bSigma^\ast}^{-1} (\by_i \by_i^\top - \bSigma^\ast) {\bSigma^\ast}^{-1}$ is also sub-exponential by Exercise 2.7.10 of \citet{vershynin2018high}.

Next, we show that each element of $n^{-1}(\bV_o^{-\frac{1}{2}})^\top \bV_o^{-\frac{1}{2}} \text{vech}(\by_i \by_i^\top - \bSigma^\ast)_{[o]} \in \mathbb{R}^{p(p-1)/2}$ is sub-exponential.
Define $\bV_d^{-\frac{1}{2}} \in \mathbb{R}^{p(p+1)/2 \times p}$ and $\bV_o^{-\frac{1}{2}} \in \mathbb{R}^{p(p+1)/2 \times p(p-1)/2}$ as sub-matrices of $\bV^{-\frac{1}{2}}$ containing $l$th column of $\bV^{-\frac{1}{2}}$ for $l \in [d]$ and for $l \in [o]$, respectively. Then, we can write
\bse
n^{-1}\bV^{-1} \text{vech}(\by_i \by_i^\top - \bSigma^\ast) = 
\begin{bmatrix}
n^{-1}(\bV_d^{-\frac{1}{2}})^\top \bV_d^{-\frac{1}{2}} & n^{-1}(\bV_d^{-\frac{1}{2}})^\top \bV_o^{-\frac{1}{2}} \\
n^{-1}(\bV_o^{-\frac{1}{2}})^\top \bV_d^{-\frac{1}{2}} & n^{-1}(\bV_o^{-\frac{1}{2}})^\top \bV_o^{-\frac{1}{2}}
\end{bmatrix}
\begin{bmatrix}
\text{vech}(\by_i \by_i^\top - \bSigma^\ast)_{[d]} \\
\text{vech}(\by_i \by_i^\top - \bSigma^\ast)_{[o]}
\end{bmatrix}.
\ese
Since we have already shown that each element of 
\bse
n^{-1}(\bV_o^{-\frac{1}{2}})^\top \bV_d^{-\frac{1}{2}} \text{vech}(\by_i \by_i^\top - \bSigma^\ast)_{[d]} + n^{-1}(\bV_o^{-\frac{1}{2}})^\top \bV_o^{-\frac{1}{2}} \text{vech}(\by_i \by_i^\top - \bSigma^\ast)_{[o]}
\ese
is sub-exponential, it is sufficient if we show that each element of $n^{-1}(\bV_o^{-\frac{1}{2}})^\top \bV_d^{-\frac{1}{2}} \text{vech}(\by_i \by_i^\top - \bSigma^\ast)_{[d]}$ is sub-exponential because the sum of two sub-exponential variables is sub-exponential. 

With some matrix algebra, we can show that 
\bse
n^{-1}(\bV_o^{-\frac{1}{2}})^\top \bV_d^{-\frac{1}{2}} \text{vech}(\by_i \by_i^\top - \bSigma^\ast)_{[d]} = \text{vech}\{{\bSigma^\ast}^{-1} \text{diag}(\by_i \by_i^\top - \bSigma^\ast) {\bSigma^\ast}^{-1}\}_{[o]}
\ese
where, for a $p \times p$ matrix $\bA$, $\text{diag}(\bA)$ is defined as a diagonal matrix whose diagonal elements are equal to that of $\bA$.
Also, the $(j,k)$th element of ${\bSigma^\ast}^{-1} \text{diag}(\by_i \by_i^\top) {\bSigma^\ast}^{-1}$ can be written as  
\begin{align*}
({\bSigma^\ast}^{-1} \text{diag}(\by_i \by_i^\top) {\bSigma^\ast}^{-1})_{jk} &= \sum_{m=1}^p {\sigma^\ast}^{jm} y_{im}^2 {\sigma^\ast}^{mk} \\ 
&= \by_i^\top \bA^{jk} \by_i
\end{align*}
with
\bse
\bA^{jk} =
\begin{bmatrix}
{\sigma^\ast}^{j1}{\sigma^\ast}^{1k} & 0 & \ldots & 0 \\
0 & {\sigma^\ast}^{j2}{\sigma^\ast}^{2k} & \ldots & 0 \\
\vdots & \vdots & \ddots & \vdots \\
0 & 0 & \ldots & {\sigma^\ast}^{jp}{\sigma^\ast}^{pk}
\end{bmatrix}\\
\ese
where ${\sigma^\ast}^{jm}$ and ${\sigma^\ast}^{mk}$ are the $(j,m)$th and the $(m,k)$th elements of ${\bSigma^\ast}^{-1}$, respectively, and $y_{im}$ is the $m$th element of $\by_i$.
Since $\by_i^\top \bA^{jk} \by_i$ is a quadratic form of a normal random vector, it follows the generalized chi-square distribution which 
can be expressed as a linear combination of independent non-central chi-square distribution \citep{imhof1961computing}
\bse
\by_i^\top \bA^{jk} \by_i = \sum_{m=1}^p \lambda_m \chi_m^2 
\ese
where $\lambda_1,\ldots,\lambda_p$ are eigenvalues of $\bA^{jk} \bSigma^\ast$. 
Also, we can easily show that the eigenvalues of $\bA^{jk} \bSigma^\ast$ are bounded because
\begin{align*}
\lambda_{\min}(\bA^{jk} \bSigma^\ast) &\geq \lambda_{\min}(\bA^{jk}) \lambda_{\min}(\bSigma^\ast)\\
&\geq \lambda_{\min}({\bSigma^\ast}^{-1} \otimes {\bSigma^\ast}^{-1}) \lambda_{\min}(\bSigma^\ast)
\end{align*}
and
\begin{align*}
\lambda_{\max}(\bA^{jk} \bSigma^\ast) &\leq \lambda_{\max}(\bA^{jk}) \lambda_{\max}(\bSigma^\ast)\\
&\leq \lambda_{\max}({\bSigma^\ast}^{-1} \otimes {\bSigma^\ast}^{-1}) \lambda_{\max}(\bSigma^\ast).
\end{align*}
Since $\sum_{m=1}^p \lambda_m \chi_m^2$ will be sub-exponential as long as $\lambda_1,\ldots,\lambda_p$ are bounded, we have shown each element of ${\bSigma^\ast}^{-1} \text{diag}(\by_i \by_i^\top) {\bSigma^\ast}^{-1}$ is sub-exponential.
Also, since the bounds for $\lambda_1,\ldots,\lambda_p$ do not depend on $j$ and $k$, the sub-exponential norm of the elements of ${\bSigma^\ast}^{-1} \text{diag}(\by_i \by_i^\top) {\bSigma^\ast}^{-1}$ is uniformly bounded.
Since $\mathbb{E}({\bSigma^\ast}^{-1} \text{diag}(\by_i \by_i^\top) {\bSigma^\ast}^{-1}) = {\bSigma^\ast}^{-1} \text{diag}(\bSigma^\ast) {\bSigma^\ast}^{-1}$, each element of ${\bSigma^\ast}^{-1} \text{diag}(\by_i \by_i^\top - \bSigma^\ast) {\bSigma^\ast}^{-1}$ is also sub-exponential by Exercise 2.7.10 of \citet{vershynin2018high}.
Hence, the elements of $n^{-1}(\bV_o^{-\frac{1}{2}})^\top \bV_d^{-\frac{1}{2}} \text{vech}(\by_i \by_i^\top - \bSigma^\ast)_{[d]}$ are sub-exponential with a uniform bound for their sub-exponential norm.
\qed


\subsection{Proof of Lemma \ref{lem:Vmax_bound}}
From \citet{cai2011constrained}, we have
\bse
\|\wh{\bOmega} - \bOmega^\ast\|_{\max} \leq 4 C_0 M^2 \sqrt{\frac{\log p}{n}}
\ese
with probability $1-4p^{-\tau}$.
Hence,
\begin{align*}
n^{-1} \|\wh{\bV}^{-1}-\bV^{-1}\|_{\max} & = \|\wh{\bOmega} \otimes \wh{\bOmega}-\bOmega^\ast \otimes \bOmega^\ast\|_{\max}\\
& = \|(\wh{\bOmega}-\bOmega^\ast+\bOmega^\ast) \otimes (\wh{\bOmega}-\bOmega^\ast+\bOmega^\ast)-(\bOmega^\ast \otimes \bOmega^\ast)\|_{\max}\\
& = \|(\wh{\bOmega}-\bOmega^\ast) \otimes (\wh{\bOmega}-\bOmega^\ast)+\bOmega^\ast \otimes (\wh{\bOmega}-\bOmega^\ast)+(\wh{\bOmega}-\bOmega^\ast) \otimes \bOmega^\ast  \|_{\max}\\
& \leq \| (\wh{\bOmega}-\bOmega^\ast) \otimes (\wh{\bOmega}-\bOmega^\ast) \|_{\max} + \| \bOmega^\ast \otimes (\wh{\bOmega}-\bOmega^\ast) \|_{\max}\\ & \quad + \| (\wh{\bOmega}-\bOmega^\ast) \otimes \bOmega^\ast \|_{\max}\\
& \leq \| \wh{\bOmega} - \bOmega^\ast \|_{\max}^2 + 2 \| \bOmega^\ast \|_{\max} \| \wh{\bOmega} - \bOmega^\ast \|_{\max}\\
& \leq 8 C_0 M^2/c \sqrt{\log p /n} 
\end{align*}
since $\| \wh{\bOmega} - \bOmega^\ast \|_{\max}^2$ is dominated by $\| \wh{\bOmega} - \bOmega^\ast \|_{\max}$ and $\| \bOmega^\ast \|_{\max} \leq c^{-1}$ by Assumption \ref{ass:eigen_bound}. \qed

\subsection{Proof of Lemma \ref{lem:re_cond_est}}
Since $\lambda_{\min}\{n^{-1} (\wh{\bV}_o^{-\frac{1}{2}})^\top \wh{\bV}_o^{-\frac{1}{2}}\} \geq \lambda_{\min}(n^{-1} \bV^{-1}) \geq C^{-2}$ by Assumption \ref{ass:eigen_bound} and $n^{-1}\| (\bV_o^{-\frac{1}{2}})^\top \bV_o^{-\frac{1}{2}}-(\wh{\bV}_o^{-\frac{1}{2}})^\top \wh{\bV}_o^{-\frac{1}{2}} \|_{\max} \leq n^{-1} \|\wh{\bV}^{-1}-\bV^{-1}\|_{\max} \leq 8 C_0 M^2/c \sqrt{\log p /n}$ by Lemma \ref{lem:Vmax_bound},
\begin{align*}
n^{-1}\balpha^\top(\wh{\bV}_o^{-\frac{1}{2}})^\top \wh{\bV}_o^{-\frac{1}{2}}\balpha &= n^{-1}\balpha^\top(\bV_o^{-\frac{1}{2}})^\top \bV_o^{-\frac{1}{2}}\balpha - n^{-1}\balpha^\top \{(\bV_o^{-\frac{1}{2}})^\top \bV_o^{-\frac{1}{2}}-(\wh{\bV}_o^{-\frac{1}{2}})^\top \wh{\bV}_o^{-\frac{1}{2}} \}\balpha\\
&\geq \frac{1}{C^2} \| \balpha \|_2^2 - n^{-1}\| (\bV_o^{-\frac{1}{2}})^\top \bV_o^{-\frac{1}{2}}-(\wh{\bV}_o^{-\frac{1}{2}})^\top \wh{\bV}_o^{-\frac{1}{2}} \|_{\max} \|\balpha\|_1^2\\
&\geq \frac{1}{C^2} \| \balpha \|_2^2 - \frac{8 C_0 M^2}{c} \sqrt{\frac{\log p}{n}} \|\balpha\|_1^2\\
&\geq \frac{1}{C^2} \| \balpha \|_2^2 - \frac{128 C_0 M^2}{c} \sqrt{\frac{\log p}{n}} \|\balpha_{\mathcal{S}}\|_1^2\\
&\geq \frac{1}{C^2} \| \balpha \|_2^2 - \frac{128 C_0 M^2}{c} s \sqrt{\frac{\log p}{n}} \|\balpha_{\mathcal{S}}\|_2^2\\
&\geq \frac{1}{C^2} \| \balpha \|_2^2 - \frac{128 C_0 M^2}{c} s \sqrt{\frac{\log p}{n}} \|\balpha\|_2^2\\
&\geq \frac{1}{C^2} \| \balpha \|_2^2 - \frac{1}{2C^2} \| \balpha \|_2^2, 
\end{align*}
where we have used
\bse
s \leq \frac{c}{256 C_0 C^2 M^2} \sqrt{\frac{n}{\log p}}.
\ese
\qed

\subsection{Proof of Lemma \ref{lem4}} \label{lem4_proof}
From \eqref{eq:supeq1} in the proof of Theorem \ref{thm2}, we have
\begin{align*}
\frac{\|\wh{\bV}_o^{-\frac{1}{2}} ({\bsigma^\ast}_{[o]} - \wh{\bsigma}_{[o]})\|_2^2}{2n} + \frac{\lambda}{2} \|\bsigma_{[o]}^\ast-\wh{\bsigma}_{[o]}\|_1 &\leq \frac{\lambda}{2} \|\bsigma_{[o]}^\ast-\wh{\bsigma}_{[o]}\|_1 + \frac{3\lambda}{2} \|(\bsigma_{[o]}^\ast-\wh{\bsigma}_{[o]})_\mathcal{S}\|_1 - \frac{\lambda}{2} \|(\bsigma_{[o]}^\ast-\wh{\bsigma}_{[o]})_{\mathcal{S}^c}\|_1\\
&= 2 \lambda \|(\bsigma_{[o]}^\ast-\wh{\bsigma}_{[o]})_\mathcal{S}\|_1.
\end{align*}
Hence, 
\begin{align*}
\lambda \|\bsigma_{[o]}^\ast-\wh{\bsigma}_{[o]}\|_1 &\leq 4 \lambda \|(\bsigma_{[o]}^\ast-\wh{\bsigma}_{[o]})_\mathcal{S}\|_1\\
&\leq 4 \lambda \sqrt{s} \|\bsigma_{[o]}^\ast-\wh{\bsigma}_{[o]}\|_2\\
&\leq 4 \lambda \sqrt{s} C^\ast \sqrt{\frac{s\log p}{n}}
\end{align*}
where the last inequality holds by Theorem \ref{thm2}.
Thus,
\bse
\|\wh{\bsigma}_{[o]} - {\bsigma^\ast}_{[o]}\|_1 = O_P \bigg( s \sqrt{\frac{\log p}{n}} \bigg).
\ese
\qed

\end{document}